\documentclass[prd,amsmath,amssymb,twocolumn,superscriptaddress,reprint,floatfix,aps]{revtex4-2}
\usepackage{graphicx}
\usepackage{bm}
\usepackage[utf8]{inputenc}
\usepackage[T1]{fontenc}
\usepackage{mathptmx}
\usepackage{etoolbox}
\usepackage{xcolor}
\usepackage[normalem]{ulem} % Normal emphasis (italics)
\usepackage{orcidlink}
\usepackage{comment}
\usepackage{hyperref}%
\hypersetup{colorlinks,%,%
 linkcolor=blue,%
 citecolor=blue,%
 urlcolor=blue
}

\newcommand{\mbx}{\mathbf{x}}

\newcommand{\mbu}{\mathbf{u}}
\newcommand{\gd}{g_{3D}}
\newcommand{\gdd}{g_{3D_2}}
\DeclareMathOperator{\sinc}{sinc}
\begin{document}

\title{Vortex Retention Mediated Turbulent Transitions in Self-Gravitating Bosonic and Axionic Condensates}

\author{Anirudh Sivakumar\orcidlink{0009-0007-9527-4555}}%
\affiliation{Department of Physics, Bharathidasan University, Tiruchirappalli 620 024, Tamil Nadu, India}

\author{Sanjay Shukla\orcidlink{0000-0001-6522-0942}}%
\affiliation{Department of Applied Physics and Science Education, Eindhoven University of Technology, 5600 MB Eindhoven, The Netherlands}

\author{Rahul Pandit}%
\affiliation{Centre for Condensed Matter Theory, Department of Physics, Indian Institute of Science, Bangalore 560012, India}

\author{Pankaj Kumar Mishra\orcidlink{0000-0003-1566-0012}}
%\email{pankaj.mishra@iitg.ac.in}
\affiliation{Department of Physics, Indian Institute of Technology Guwahati, Guwahati 781039, Assam, India}

\author{Paulsamy Muruganandam\orcidlink{0000-0002-3246-3566}}
%\email{anand@bdu.ac.in}
\affiliation{Department of Physics, Bharathidasan University, Tiruchirappalli 620 024, Tamil Nadu, India}
\affiliation{Department of Medical Physics, Bharathidasan University, Tiruchirappalli 620 024, Tamil Nadu, India}

\begin{abstract}
We investigate turbulent spin-down dynamics in self-gravitating Bose-Einstein condensates, comparing purely bosonic and axionic (higher-order interacting) systems. Through simulations of the Gross-Pitaevskii-Poisson system, we study condensates pinned to a crust potential undergoing rapid rotation slowdown. We find that axionic condensates exhibit more uniform density profiles and smaller sizes compared to their bosonic counterparts for similar interaction strengths, which facilitates earlier vortex entry. The sudden spin-down triggers vortex depinning and a turbulent cascade. For comparable sizes, both systems exhibit a short-lived Kolmogorov energy cascade (\(k^{-5/3}\) scaling) followed by a transition to Vinen turbulence (\(k^{-1}\) scaling). Crucially, their responses diverge with increasing interaction strength (and thus condensate size): the axionic system increasingly deviates from Kolmogorov scaling because of enhanced vortex retention, a trend quantitatively confirmed by analyzing the vortex fraction and its dependence on the final rotation frequency. Spectral analysis reveals that the growth of incompressible energy is primarily driven by quantum pressure during vortex detachment, rather than by compressible flows. The compressible spectrum shows thermalization (\(k\) scaling). Our results demonstrate how distinct nonlinearities govern vortex dynamics and turbulent dissipation in self-gravitating quantum fluids.
\end{abstract}

\date{\today}
\maketitle

\section{Introduction}
\label{sec:intro}

Turbulence in quantum fluids, initially conceptualized theoretically by Feynman~\cite{Feynman1955} and subsequently realized experimentally by Vinen in 1957~\cite{Vinen1957}, provides a setting in which macroscopic hydrodynamic behaviour emerges from underlying quantum features~\cite{paoletti2011quantum,Barenghi:CUP2023,Tsubota:OUP2025}. In contrast to classical turbulence, quantum turbulence (QT) consists of disordered tangles of quantized vortices-topological defects whose circulation is quantized in units of \(h/m\)~\cite{White2014}. While early studies focused on counterflowing superfluid helium~\cite{Guo2010}, QT has more recently been observed in Bose–-Einstein condensates (BECs), which offer a highly controllable and compressible platform with tunable, weak interatomic interactions~\cite{Tsubota:OUP2025, barenghi2023types, White2014, Henn2009}. Similar to turbulence in classical fluids, QT often proceeds through a cascade of energy from large to small length scales, bearing striking phenomenological similarities to the classical Richardson cascade, despite the discrete nature of the vorticity field \citep{Barenghi2014, Barenghi:CUP2023}. This cascade manifests itself as a Kolmogorov-type power-law scaling of the energy spectrum $E(k) \sim k^{-5/3}$, as a function of the wavenumber $k$, for the condensate often referred to as Kolmogorov or K41 scaling~\cite{frisch1995turbulence}). This scaling has been observed in three-dimensional (3D) superfluids~\cite{Maurer1998, Nore1997, Araki2002} and also in BECs~\cite{Kobayashi2007, Kobayashi2008, Neely2013, Wilson2013}, displaying a direct~\cite{Numasato2010} and inverse energy cascades~\cite{Mueller2020, Reeves2013}. Apart from the Kolmogorov regime, the Vinen (ultraquantum) regime  has also been identified, which describes random distributions of vortices in superfluids and BECs~\cite{Volovik2004, Bradley2006, Cidrim2017}. In this regime, the energy spectra of the condensates display a $k^{-1}$ scaling in an inertial range along with a temporal decay of vortex-line density as $\sim t^{-1}$~\cite{Cidrim2017}.

Quantum turbulence provides a valuable framework for probing universal features of turbulence dynamics across different physical systems~\citep{Tsubota2013, Barenghi:CUP2023, Tsubota:OUP2025}. In recent years, researchers have identified novel turbulent regimes, including strong quantum turbulence~\cite{barenghi2023types, MiddletonSpencer2023, Navon2016, Sivakumar2024a} and rotating turbulence~\cite{AmetteEstrada2022, Estrada2022, Sivakumar2024}. Moreover, the conceptual and computational framework of QT has been extended beyond laboratory atomic gases to describe complex astrophysical and cosmological systems. Examples include the interior of neutron stars, where a superfluid neutron component coexists with a crystalline crust \citep{Haskell2015}, and dark matter halos, which may consist of ultra-light axion-like particles forming the self-gravitating condensates~\citep{Hui2017, Chavanis2012, Chavanis2014, Ferreira2021, Boehmer2007, shukla_2025_GPPE_turbulence}. In such self-gravitating quantum systems, distinctive spectral signatures have been reported: the energy spectra may exhibit a $\sim k^{-1.1}$ scaling, reminiscent of counterflow turbulence in superfluids~\cite{Mocz2017}, whereas a classical Kolmogorov scaling emerges when condensates collide at sufficiently high velocities~\cite{Sivakumar2025}.

In neutron stars, the coupling between the superfluid core and the solid crust is thought to play a central role in the occurrence of \textit{pulsar glitches} --sudden spin-up events observed in pulsar rotation rates~\cite{Radhakrishnan1969, Boynton1969}. These events arise from the transfer of the angular momentum from the superfluid component to the crust via quantized vortices~\cite{Manchester2017}. Within the vortex-avalanche framework, the gradual electromagnetic spin-down of the star leads to vortex pinning at nuclear lattice sites in the crust~\cite{Jones1986, Khomenko2018, Howitt2022}. Pinning is most effective when the characteristic length scale of the pinning sites is comparable to the superfluid healing length~\cite{Tsubota1993}, a condition also encountered in atomtronic~\cite{Amico2021} and superconducting circuits~\cite{Nelson1993}. As the rotational lag between the crust and the superfluid increases, a critical stress is reached, resulting in collective vortex unpinning and outward motion. The associated angular momentum transfer produces a glitch~\cite{Melatos2008}. Depinning occurs when the Magnus flow around pinning sites exceeds a critical velocity~\cite{Loennborn2019, Sonin1997, Stockdale2021, Schwarz1981}, with the subsequent dynamics determined by the height and spatial extent of the pinning potential~\cite{Liu2024}.

Over the years, the modelling of glitch phenomena in neutron stars as the spin-down of a self-gravitating, pinned superfluid BEC within the GP formalism has provided significant insights~\cite{Verma2022, Warszawski2011, Warszawski2012, Melatos2015}. Such models have resolved the associated vortex dynamics and have shown, for instance, that large vortex populations can reduce individual glitch sizes, provided that there is an observational upper bound~\cite{Warszawski2011}. In these simulations, the pinning effect is typically realised using a crust potential, analogous to the interface between the inner and outer core of a neutron star~\cite{Verma2022, Shukla2024, Lattimer2004}. More recently, Liu, Baggaley, Barenghi, and Wood~\cite{Liu2025} have demonstrated that, for sufficiently small spin-downs, glitches, which are consistent with a vortex-avalanche mechanism, can occur through collective vortex motion; and Verma, Shukla, Brachet, and Pandit~\cite{Verma2022,Shukla2024} have investigated glitch phenomena in gravitationally collapsed bosonic stars in the presence of a crust and magnetic field using the Gross-Pitaevskii-Poisson framework~\cite{Nikolaieva2023}.

Concurrently, there is growing interest in \textit{axionic} or \textit{higher-order} interacting BECs as minimal models for fuzzy dark matter. These systems are described by a modified GP equation that includes an additional quintic nonlinear term, representing three-body interactions or effective potentials in certain cosmological contexts \citep{Chavanis2014, Schiappacasse2018}. This axionic term qualitatively alters the condensate's static and dynamic properties, leading to distinct density profiles, stability regimes, and collective excitation spectra compared to standard two-body interacting (bosonic) BECs. A critical open question is how this different underlying microphysics---bosonic versus axionic interactions---influences the macroscopic turbulent dynamics during non-equilibrium processes, such as rapid spin-down. Do both systems follow the same route to turbulence, or do their distinct nonlinearities lead to fundamentally different dissipation pathways?. Additionally, recent studies have shown that QT dampens collective excitation modes in BECs~\cite{Lee2025, Ferrand2021}. As these modes are responsible for vortex avalanches and glitches, quantifying the onset of turbulence in such systems is crucial; see, e.g., the recent review on different models of the GP equation from microscopic to astrophysical length scales~\cite{shukla2025selfgravitatingsuperfluidsgrosspitaevskiipoissonframework} and references therein.

In this work, we perform a direct numerical comparison of turbulent spin-down in self-gravitating bosonic and axionic BECs pinned to an external crust potential. Our objectives are threefold: (1) to characterise and contrast how bosonic ($\gd$) and axionic ($\gdd$) interactions affect the static condensate profile and the critical frequency for vortex nucleation; (2) to simulate an abrupt spin-down and analyse the resulting vortex depinning dynamics, the generation of compressible and incompressible flows, and the inter-component energy exchange---particularly the role of quantum pressure in driving incompressible flow; and (3) to decipher the signatures of turbulence in the kinetic energy spectra of both systems, identifying universal features like the Kolmogorov cascade and system-specific deviations. A key focus is to understand how vortex retention---quantified through vortex population decay and its dependence on the final rotation frequency---differs between the two systems and governs the stability of the turbulent cascade. This three-dimensional, self-gravitating geometry provides a distinct physical context for generating quantum turbulence, complementing studies of spin-down in two-dimensional condensates within hard-walled traps~\cite{Sivakumar2026}.

We solve the coupled Gross-Pitaevskii-Poisson (GPP) system of equations numerically; we incorporate both self-gravity and a pinned crust potential. Starting from a rapidly rotating equilibrium state, we impose a linear ramp-down of the rotation frequency. We find that, while both bosonic and axionic systems exhibit vortex depinning and a transition to turbulence, the axionic condensate’s compact and non-uniform structure allows vortices to nucleate more easily than in the bosonic system. The turbulent state in both cases shows a  limited inertial range with a Kolmogorov-like ($k^{-5/3}$) scaling in the incompressible kinetic energy spectrum, followed by a ($k^{-1}$) scaling indicative of a sparse vortex network (Vinen turbulence). However, a key difference emerges as the interaction strength increases: the axionic system favours vortex retention and exhibits a pronounced deviation from K41 scaling and a suppression of the Kolmogorov cascade, whereas the bosonic system continues to show the classical K41 cascade. Furthermore, the compressible spectrum reveals signatures of  thermalisation, modulated by the condensate’s finite size.

The outline of our manuscript is as follows. In Sec.~\ref{sec:model}, we introduce the mean-field model that describes the dynamics of self-gravitating and axionic condensates. This section also includes detailed calculations of key quantities, such as the various components of the energy and relevant length scales within our model. In Sec.~\ref{sec:analysis}, we analyze the phenomena of vortex depinning and retention, and examine their impact on the energy cascade for self-gravitating bosonic and axionic condensates. Finally, in Sec.~\ref{sec:summary}, we summarize our principal findings, discuss their implications, and outline potential directions for future work.

\section{Model and Methods}
\label{sec:model}
We consider a three-dimensional condensate whose dynamics are described by the dimensionless Gross--Pitaevskii--Poisson system of equations \cite{Shukla2024Axion}:
\begin{subequations}\label{eq:gpe:poisson}
\begin{align}
  \mathrm{i}\,\frac{\partial \psi}{\partial t} & = \left(-\frac{1}{2}\nabla^2 + V_\mathrm{conf} + V_\mathrm{int} - \Omega(t)\,L_z\right)\psi, \label{eq:gpe}\\
  \nabla^2 \Phi & = \lvert \psi \rvert^2. \label{eq:poisson}
\end{align}
\end{subequations}
Here,
\(V_\mathrm{int} = \gd \lvert \psi \rvert^2 + \gdd \lvert \psi \rvert^4\) and
\(V_\mathrm{conf} = G \Phi + \frac{1}{2}\Omega(t)\, r_\perp^2 + V_\mathrm{crust}\),
where $\psi \equiv \psi(\mathbf{r},t)$ is the condensate wavefunction with $\mathbf{r} \equiv (x,y,z)$. The operator $\nabla^2$ denotes the three-dimensional Laplacian, defined as $\nabla^2 = \partial_x^2 + \partial_y^2 + \partial_z^2$.
We obtain this dimensionless form by considering the following transformations $\mathbf{r}\to \mathbf{r}/L$, $t \to t / T$, $E\to E/\epsilon$ of length, time, and energy, respectively. We take the scaling factors as given in \cite{Shukla2024Axion}. The condensate is confined by the confining potential of the form $V_{\mathrm{conf}}$, which contains the gravitational potential $\Phi\equiv \Phi(\mathbf{r})$ obtained by solving the density-dependent Poisson equation Eq.~\eqref{eq:poisson} with $G$ being the gravitational strength. The confinement is also provided by the centrifugal potential of the form $\frac{1}{2}\Omega r_\perp^2$, where $\Omega$ is the time-dependent rotation frequency and $r_\perp = \sqrt{x^2 + y^2}$ is the radial width of the condensate. The centrifugal potential ensures the uniform distribution of density under the effects of rotation.  The condensate interaction is modeled by the potential $V_\mathrm{int}$, which consists of a standard term with strength $\gd = {4\pi \hbar^2 a}/{m}$ and a higher-order term with strength $\gdd = {32\pi^2 \hbar^4 a^2}/{(3m^3c^2)}$. Here $a$, $m$, and $c$ are the $s$-wave scattering length, particle mass, and speed of light, respectively. We incorporate these terms in the non-dimensional equation by taking $\hbar = m = 1$. For our axionic system, we consider an attractive scattering length ($a < 0$) for the bosonic component, which is counteracted by a repulsive quintic term $\gdd$ to prevent the condensate from collapsing. The condensate is placed in a rotating frame, described by the term $\Omega(t) L_z$, where $L_z = -\mathrm{i}(x\partial_y - y\partial_x)$ is the $z$-component of the angular momentum. The time-dependent rotation frequency $\Omega(t)$ describes the rotational deceleration profile of the condensate as follows:
\begin{align}
    \Omega(t) = 
    \begin{cases}
        (\Omega_0 - \Omega_f)\cos^2\left(
        \frac{\pi }{2t_s}\right) + \Omega_f, & \text{if }t\leq t_s, \\
        \Omega_f,& \text{if } t > t_s,
    \end{cases}
    \label{eq:rotfreq}
\end{align}
where $\Omega_0$ and $\Omega_f$ are the initial and final rotation frequencies of the condensate and $t_s$ is the spin-down time. 

To study the effects of depinning in condensates, we include a crust potential, \( V_\mathrm{crust} \), in the confining term. This potential, given by \cite{Verma2022, Shukla2024}, is defined as
\begin{align}
 V_\mathrm{crust}(\mathbf{r},t) = V_0c\exp\Big(-\frac{(\lvert\mathbf{r}_p\rvert-r_\mathrm{crust})^2}{(\Delta r_\mathrm{crust})^2}\Big)\tilde{V}(x_\theta,y_\theta),
 \end{align}
where \(\tilde{V}(x_\theta,y_\theta) = 3 + \cos(n_\mathrm{crust}x_\theta) + \cos(n_\mathrm{crust}y_\theta)\) and the rotated coordinates are \(x_\theta = x_p\cos \theta + y_p\sin\theta \) and \(y_\theta = -x_p\sin \theta + y_p\cos\theta \). Here, \(n_\mathrm{crust}\) determines the number of pinning sites, \(r_\mathrm{crust}\) is the crust radius, and \(\Delta r_\mathrm{crust}\) is the crust thickness. The parameters for this potential are based on Refs. \cite{Shukla2024, Verma2022}. It is worth noting that Refs.~\cite{Verma2022,Shukla2024} employed a dynamic-crust model by incorporating the evolution of $\theta$ with time along with a frictional coupling of the rotating condensate that facilitated the transfer of angular momentum. In contrast, in the present work, the crust is static, and the effective friction arises as an effect of the controlled ramp-down of the rotation frequency $\Omega$.
%Ref.~\cite{Verma2022} considers a dynamic crust model, where $\theta$ evolves with time and provides a frictional effect on the rotating condensate causing a transfer of angular momentum. In our case, the crust is static and the frictional effect is induced by the ramping-down of rotation frequency $\Omega$. 

To characterize the turbulence and energy cascades resulting from vortex dynamics, we conduct a detailed analysis of the kinetic energy spectra across various length and time scales. We decompose the kinetic energy into compressible and incompressible components~\cite{Nore1997} by transforming the incompressible and compressible velocity fields into $k$ space integrals using Parseval's theorem. We employ the numerical implementation developed by Bradley, Kumar, Pal, and Yu~\cite{Bradley2022}, which evaluates these integrals using the angle-averaged Wiener-Khinchin theorem, as illustrated below.
\begin{align}
  \varepsilon_{\mathrm{kin}}^\zeta (k) = \frac{m}{2}\int d^d\mathbf{x} \, \Lambda_d\left(k, \lvert \mathbf{x} \rvert\right)\, C\left[\mathbf{u}^\zeta, \mathbf{u}^\zeta\right](\mathbf{x}), \label{eq:wkspec}
\end{align}
where $\varepsilon_{\mathrm{kin}}^\zeta(k)$ is the angle-averaged kinetic energy spectrum for a given component $\zeta\in \{i,c,q\}$, where $i$, $c$, and $q$ denote incompressible, compressible, and quantum, $d$ specifies the dimension, $C[\mbu^\zeta,\mbu^\zeta](\mbx)$ represents the two-point auto-correlation function in position for a given component of the velocity field $\mbu$ and $\Lambda_d(k,\lvert\mathbf{x}\rvert)$ is the dimension-dependent kernel function:
\begin{align}
  \Lambda_{d}(k,r) = 
  \begin{cases}
    \frac{1}{2\pi}k J_0(kr), & \mbox{for } d=2\,; \\
    \frac{1}{2\pi^2}k^2 \sinc(kr), & \mbox{for } d=3\,.\\
  \end{cases}
  \label{eq:kernel}
\end{align}
Equation~\eqref{eq:wkspec} implies that, for any position-space field, there exists a spectral density equivalent to an angle-averaged two-point correlation in $k$ space. New vector velocity fields $\mathbf{w}^{i,c,q} = (i\mbu^i,i\mbu^c,i\mbu^q)e^{i\phi}$ are introduced. These fields do not satisfy the irrotational or solenoidal conditions because of the quantum phase $\phi$. By computing the vector convolution of these fields, the kinetic spectral density can be expressed as
\begin{align}
    \varepsilon_\mathrm{kin}(k)  = & \varepsilon^i_\mathrm{kin}(k) + \varepsilon^c_\mathrm{kin}(k) + \varepsilon^q_\mathrm{kin}(k) \notag \\ &  + \varepsilon^{iq}_\mathrm{kin}(k) + 
    \varepsilon^{cq}_\mathrm{kin}(k) + 
    \varepsilon^{ic}_\mathrm{kin}(k) .
\end{align}
In addition to the energy contributions from individual velocity components $\varepsilon^\zeta_\mathrm{kin}(k)$, there are contributions from exchange or coupling terms $\varepsilon^{\zeta\zeta^{'}}_\mathrm{kin}(k)$ to the total energy density, where $\zeta,\zeta^{'}\in \{i,c,q\}$. Although these exchange terms do not contribute to the total kinetic energy of the condensate, they describe the distribution and flow of energy among the components. 

To characterize the energy cascades, we consider several length scales: the Thomas-Fermi radius $R_{\mathrm{TF}}$, the inter-vortex separation $\ell_0$, the coherence length $\xi_{c}$, and the healing length $\xi$. The intervortex separation $\ell_0$ is given by $\ell_0 = (V / L_{v})^{\frac{1}{2}}$. Here $V$ is the condensate volume and $L_{v}$ is the vortex length given by 
\begin{align}
  L_v = 2\pi \frac{\int_{k_{\mathrm{mn}}}^{k_{\mathrm{mx}}} n(k) dk}{\int_{k_{\mathrm{mn}}}^{k_{\mathrm{mx}}} n_{s}(k) dk}\,,
\end{align}
as shown in \cite{AmetteEstrada2022}, with $n(k)$ the momentum spectrum of the condensate and $n_s(k)$ the momentum spectrum of a single-vortex condensate. $k_{\mathrm{mn}}$ is the minimum value from which we integrate the spectra; we choose $k_{\mathrm{mn}} = 10$; and $k_{\mathrm{mx}}$ is the maximum value in the $k$ array. For the kinetic-energy spectra, the corresponding Fourier-space wavenumbers $k_{TF} = 2\pi / R_{TF}$, $k_{\ell_0} = 2 \pi / \ell_0$, and $k_\xi = 2\pi / \xi$ are employed.

\section{Vortex Depinning, Retention, and Turbulent Cascades in Bosonic and Axionic condensates}
\label{sec:analysis}

In this section we investigate the dynamics of vortex depinning and retention to uncover their roles in the turbulent energy cascade for bosonic and axionic condensates. The depinning of vortices from pinning sites, which can occur because of external forces or changes in system parameters, is critical for understanding the onset of turbulence and the energy transfer process~\cite{Melatos2008, Sonin1997}. In self-gravitating systems, such as neutron stars and axionic condensates, vortex retention and the subsequent energy cascades are strongly influenced by the system's nonlinearity and interactions with the gravitational and bosonic or axionic fields~\cite{Hui2017, Haskell2015}. By analyzing these processes, we elucidate the connection between vortex dynamics and the macroscopic behavior of the turbulent cascade; and we compare systematically their manifestations in both bosonic and axionic condensates~\cite{Navon2016, MiddletonSpencer2023}.

The simulation is performed by numerically solving the modified Gross-Pitaevskii-Poisson system of equations, \eqref{eq:gpe} and \eqref{eq:poisson}, using the split-step Crank-Nicolson method \cite{Muruganandam2009, Vudragovic2012, YoungS2017, Kumar2019}. A detailed description of the numerical implementation is provided in Appendix \ref{sec:appendix:a}. The computational domain is a $256 \times 256 \times 256$ grid with spatial steps $dx = dy = dz = 0.075$ and a time step $dt = 5 \times 10^{-4}$.  To verify numerical convergence, additional simulations were performed on a finer grid size $512 \times 512 \times 512$ with correspondingly reduced spatial and temporal step sizes. All the relevant observation pertaining to both the axionic and bosonic cases remain quantitatively consistent, confirming the robustness of our numerical simulations runs; of course, the simulation on grid size $512^3$ requires a significantly higher computational infrastructure than that for the $256^3$ grid sizes. In this paper we mainly focus to analyze the effects of the bosonic ($\gd$) and axionic ($\gdd$) interaction terms on the turbulence, depinning, and the presence of the vortex expulsion. To this end, we perform spin-down simulations of a condensate with and without axionic interactions.

The initial rotation frequency is set to $\Omega_0 = 8.0$, and the condensate is spun down to a final frequency $\Omega_\mathrm{f} = 1.0$. The gravitational strength is $G = 100$. For the axionic condensate, we fix $\gd = -15$, which corresponds to the global minimum of the axionic system \cite{Shukla2024Axion}, and vary $\gdd$ to achieve different condensate sizes. For the non-axionic (bosonic) condensate, we set $\gdd = 0$ and vary $\gd$ in a similar manner. It is important to note that an axionic condensate and a bosonic condensate, with numerically identical interaction parameters (i.e., $\gdd = \gd$), are not directly comparable given their distinct physical origins. Therefore, our analysis is restricted to studying how an increase in $\gdd$ and $\gd$ affects the spin-down mechanics, rather than comparing the exact values of these interaction strengths.

\begin{figure}[ht!]
    \centering
    \includegraphics[width=\linewidth]{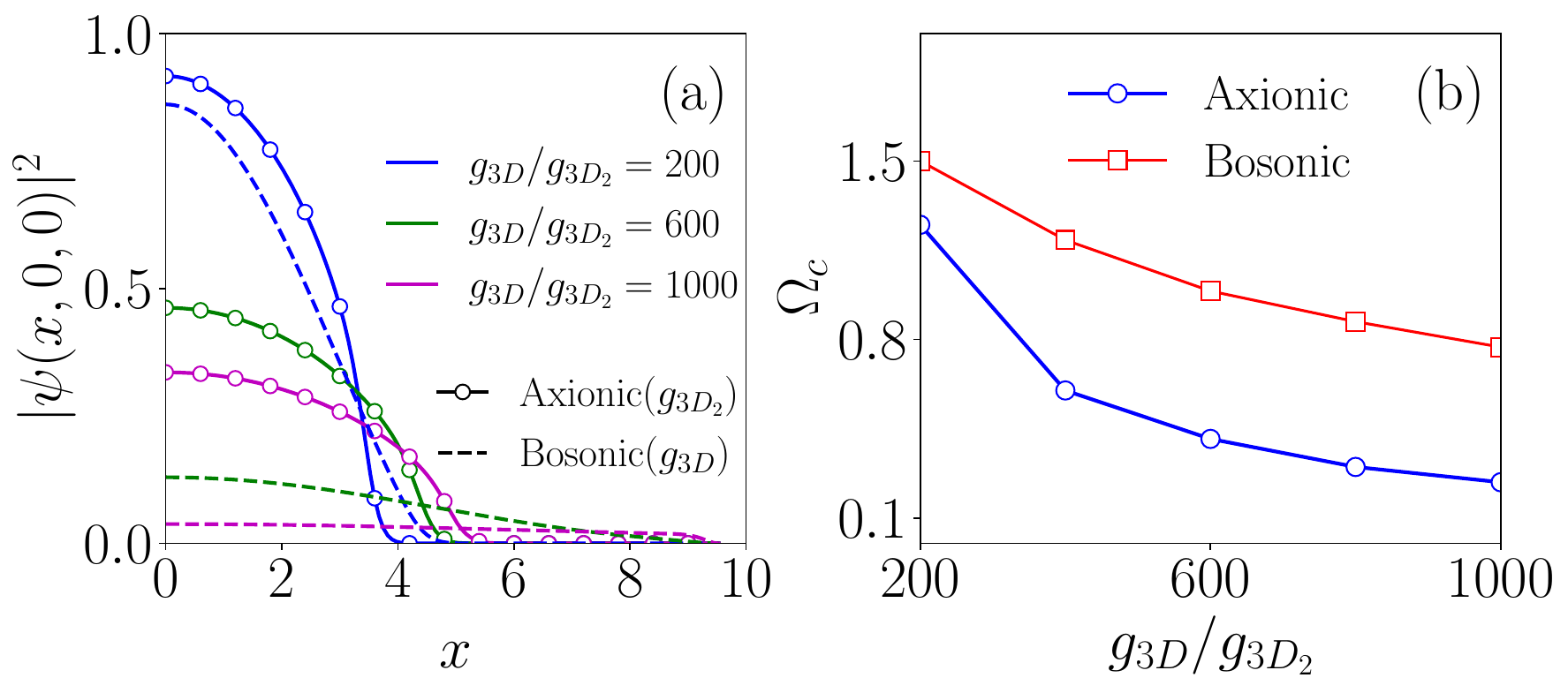}
    \caption{(a) One-dimensional density profiles of the bosonic and axionic condensates for different interaction strengths, illustrating the enhanced spatial localization induced by higher-order axionic nonlinearity. (b) Critical rotation frequency $\Omega_c$ as a function of nonlinearity strength for bosonic and axionic condensates, showing that vortices nucleate more readily in the axionic case than in the bosonic system.}
    %\caption{(a)One-dimensional density profiles of the bosonic and axionic condensates for different interaction strengths, illustrating the enhanced spatial localization induced by higher-order axionic nonlinearity. (b) Critical rotation frequency $\Omega_c$ as a function of nonlinearity strength for bosonic and axionic condensates, showing that vortices nucleate more readily in the axionic case.}
    \label{fig:1dprof}
\end{figure}

In the absence of a crust potential and rotation, we analyze the one-dimensional density profiles of bosonic ($\gdd=0$) and axionic ($\gd=-15$) condensates for different interaction values of the two-body and higher-order interaction strengths $\gd$ and $\gdd$, respectively, as shown in  Fig.~\ref{fig:1dprof}(a). %

The density profiles show that the higher-order axionic interaction enhances spatial nonuniformity, rendering the axionic condensate more compact than its bosonic counterpart for comparable interaction strengths. Moreover, increasing the higher-order interaction $\gdd$ leads to a slower expansion of the condensate compared to the growth induced by the two-body interaction $\gd$. For sufficiently small values of $\gd$ and $\gdd$, the density profiles of the two systems are nearly indistinguishable. Figure~\ref{fig:1dprof}(b) shows that the critical rotation frequency $\Omega_c$, defined as the threshold above which vortices nucleate, decreases with increasing interaction strength in both bosonic and axionic condensates. Notably, vortices nucleate more readily in the axionic condensate than in the bosonic case, despite the latter having a larger spatial extent.

%the condensate profiles are similar. Fig.~\ref{fig:1dprof}(b) shows that the critical rotation frequency ($\Omega_c$), which is the threshold frequency above which the condensate is threaded by vortices, decreases with respect to bosonic/axionic interactions. Also, it can be inferred that vortices enter the axionic condensate easily compared to the bosonic case, despite the relatively larger size of the bosonic condensate.

We perform the spin-down by initially preparing the condensates using the imaginary time scheme within a crust potential, rotating at frequency $\Omega_0$. The rotation frequency is ramped down according to Eq.~\ref{eq:rotfreq} until a final rotation frequency $\Omega_\mathrm{f}$ is reached. This protocol mimics pulsar spin-downs in Gross-Pitaevskii analogs, where abrupt changes in angular velocity can trigger vortex avalanches, widely believed to underlie glitch events in neutron stars~\cite{Warszawski2011, Warszawski2012, Liu2025, Loennborn2019}. To capture this effect, we employ a spin-down rate $\dot{\Omega} = 3.0$, corresponding to a spin-down time $t_s \simeq 2.3$, fast enough to drive non-equilibrium dynamics but slow enough to avoid numerical instabilities.
%%%%%%%%%%%%%%%%%%%%%%%%%%%%%%%%%%%%%%%%%%%%%%%%%%%%%%%%
\begin{figure*}
    \centering
    \includegraphics[width=\linewidth]{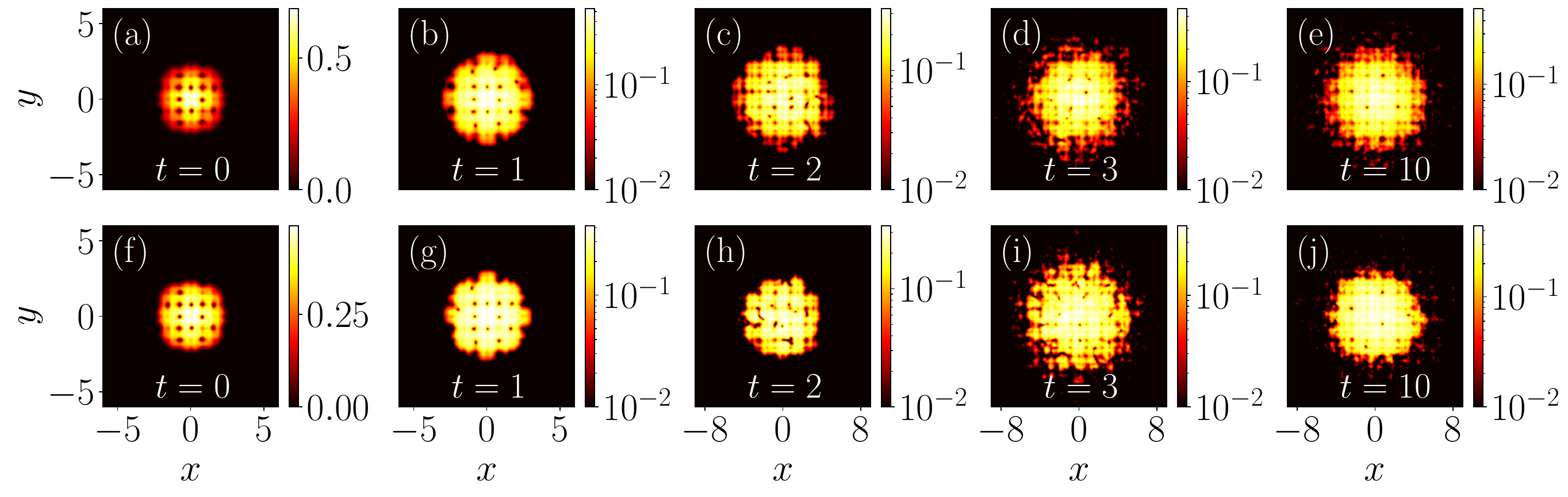}
    \caption{Snapshots of the log-normalized condensate density in the $x-y$ plane  for (a)-(e) bosonic case ($\gd =  200$) and (f)-(j) axionic case ($\gdd = 200$) at various times during the spin-down process. In both systems, vortices and compressible density excitations are expelled toward the condensate periphery as spin-down proceeds. The color bars indicate the density $|\psi|^2$.}
    \label{fig:densityxy}
\end{figure*}
%%%%%%%%%%%%%%%%%%%%%%%%%%%%%%%%%%%%%%%%%%%%%%%%%%%%%%%%
In Fig.~\ref{fig:densityxy}, we show the density profiles for both bosonic and axionic cases, which reveal the interplay between centrifugal, self-gravitating, and interaction potentials. Initially, the centrifugal potential dominates over the self-gravitating potential because of the high rotation frequency. As the rotation slows, the self-gravitating and interacting potentials become increasingly important, leading to a more compact and uniform density distribution. Furthermore, the crust potential influences the dynamics by facilitating vortex depinning: the Magnus flow induced by the spin-down reaches a critical velocity, liberating vortices from pinning sites and generating density fluctuations~\cite{Liu2025, Liu2024}. These fluctuations contribute primarily to the compressible component of the flow, which is largely confined to the periphery of the condensate  [Figs.~\ref{fig:densityxy}(d) and (j)], a behavior reminiscent of that in a bosonic condensate~\cite{Sivakumar2025}. This contrasts with standard atomic BECs, where compressible excitations are reflected by the trap boundaries, remaining trapped within the condensate~\cite{Sivakumar2026}. 

%The presence of the crust accelerates vortex decay due to a depinning mechanism, as discussed by Liu and colleagues~\cite{Liu2025, Liu2024}, who attribute it to Magnus flow around the pinning sites of the crust. 

%The Magnus flow, induced by spin-down of the condensate, reaches a critical velocity that triggers vortex depinning from the pinning sites defined by the crust potential. This depinning generates density fluctuations, which contribute to the compressible flow. As shown in the density snapshots [Fig.~\ref{fig:densityxy}(d) \& (j)], this compressible flow is primarily confined to the condensate periphery---a behavior also reported in self-gravitating systems~\cite{Sivakumar2025}. This contrasts with compressible flow in standard atomic BECs, where the hard walls from a box or harmonic trap reflect compressible waves, trapping them within the condensate~\cite{Sivakumar2026}. 

%The sudden spin-down and the associated vortex depinning and compressible fluctuations generate a turbulent flow, characterized by the energy spectrum of the incompressible and compressible components of the kinetic energy. 

\begin{figure}[!ht]
\includegraphics[width=\linewidth]{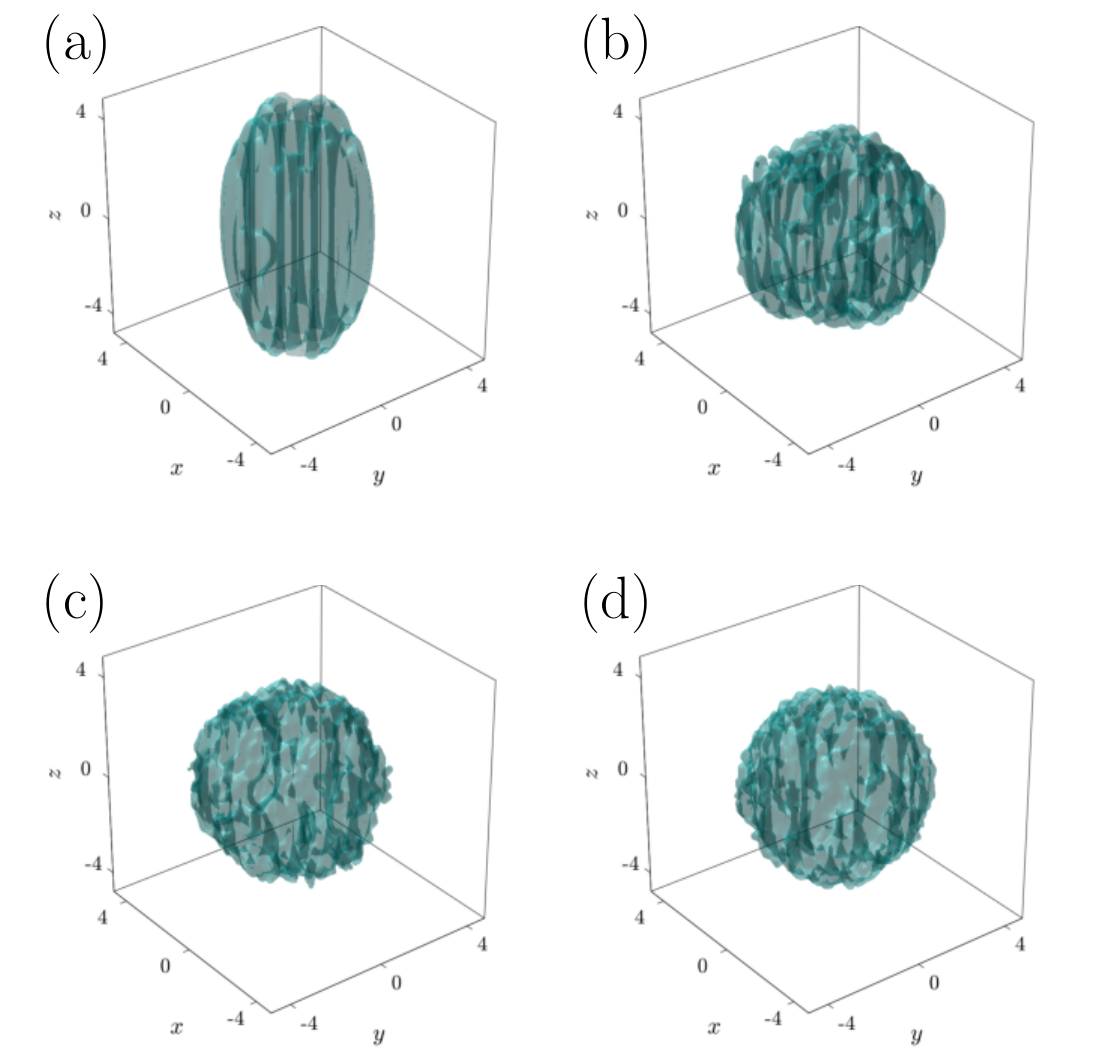}
\caption{Three-dimensional density isosurfaces of the axionic condensate with $\gdd = 200$ undergoing spin-down in the presence of a crust potential at times (a) $t = 1$, (b) $t = 2$, (c) $t = 5$, and (d) $t = 10$. Quantized vortices appear as thin tubular density depletion. }
\label{fig:3d:density}
\end{figure}

The combined effect of sudden spin-down, vortex depinning, and compressible excitations drives the system into a turbulent state. This turbulence is characterized by the redistribution of energy between incompressible and compressible components of the kinetic energy. To capture the full three-dimensional evolution, we visualize the condensate using density isosurfaces. Figure~\ref{fig:3d:density} shows the axionic condensate with $\gdd = 200$ undergoing spin-down in the presence of a crust. The snapshots clearly illustrate the transition from a centrifugally distorted, vortex-laden configuration to a more compact, self-gravitating state, highlighting how vortex dynamics mediates the redistribution of energy and the emergence of turbulence.

%Following the analysis of the 2D density profiles for both bosonic and axionic condensates [Fig.~\ref{fig:densityxy}], we examine the three-dimensional structure to better visualize the condensate’s evolution during spin-down. Figure \ref{fig:3d:density} depicts three-dimensional density isosurfaces of the axionic condensate with $\gdd = 200$ undergoing spin-down in the presence of a crust potential at various instances. The snapshots reveal the dynamical transition from a centrifugally distorted, vortex-rich state to a more compact, gravitationally dominated configuration.
%%%%%%%%%%%%%%%%%%%%%%%%%%%%%%%%%%%%%%%%%%%%%%%%%%%%%%%%
%%%%%%%%%%%%%%%%%%%%%%%%%%%%%%%%%%%%%%%%%%%%%%%%%%%%%%%%
At $t=1$ [Fig.~\ref{fig:3d:density}(a)], when the condensate rotates rapidly ($\Omega\simeq \Omega_0$), centrifugal forces elongate it along the rotation axis ($z-$ direction), and multiple quantized vortices appear as thin, tube-like density depressions threading the superfluid. As spin-down progresses at $t=2$ [Fig.~\ref{fig:3d:density}(b)], the weakening centrifugal support causes radial contraction of the condensate. By $t=5$ [Fig.~\ref{fig:3d:density}(c)], near the end of the spin-down ramp, self-gravitating and mean-field interactions dominate, producing a more spherical central density. Finally, at $t=10$ [Fig.~\ref{fig:3d:density}(d)], after reaching the final rotation frequency $\Omega_f$, the condensate settles into a nearly axisymmetric, vortex-depleted state, with remaining vortices expelled or pinned at the periphery. This 3D visualization confirms that vortex expulsion and the transition from rotation-dominated to gravity-dominated equilibrium, inferred from 2D density profiles, occur coherently across the full condensate volume.

%At $t = 1$ [Fig.~\ref{fig:3d:density}(a)], during rapid rotation ($\Omega \approx \Omega_0$), the condensate gets elongated along the rotation axis (the $z$-direction) due to centrifugal forces, and multiple quantized vortices (visible as thin, tube-like density depressions) thread the superfluid. As spin-down proceeds at $t = 2$ [Fig.~\ref{fig:3d:density}(b)], the reduction in rotation weakens the centrifugal support, causing the condensate to contract radially. By $t = 5$ [Fig.~\ref{fig:3d:density}(c)], nearing the end of the spin-down ramp, the self-gravitating and mean-field potentials begin to dominate, leading to a more spherical central density concentration. Finally, at $t = 10$ [Fig.~\ref{fig:3d:density}(d)], well after reaching the final rotation frequency $\Omega_f$, the system settles into a nearly axisymmetric, vortex-depleted state where the remaining vortices are either expelled or pinned at the periphery. This 3D visualization confirms that the vortex expulsion and the transition from rotation-dominated to gravity-dominated equilibrium, inferred from 2D cuts, occur coherently throughout the full volume of the condensate.

%%%%%%%%%%%%%%%%%%%%%%%%%%%%%%%%%%%%%%%%%%%%%%%%%%%%%%%%%%%%%%%
\begin{figure}[ht!]
    \centering
    \includegraphics[width=\linewidth]{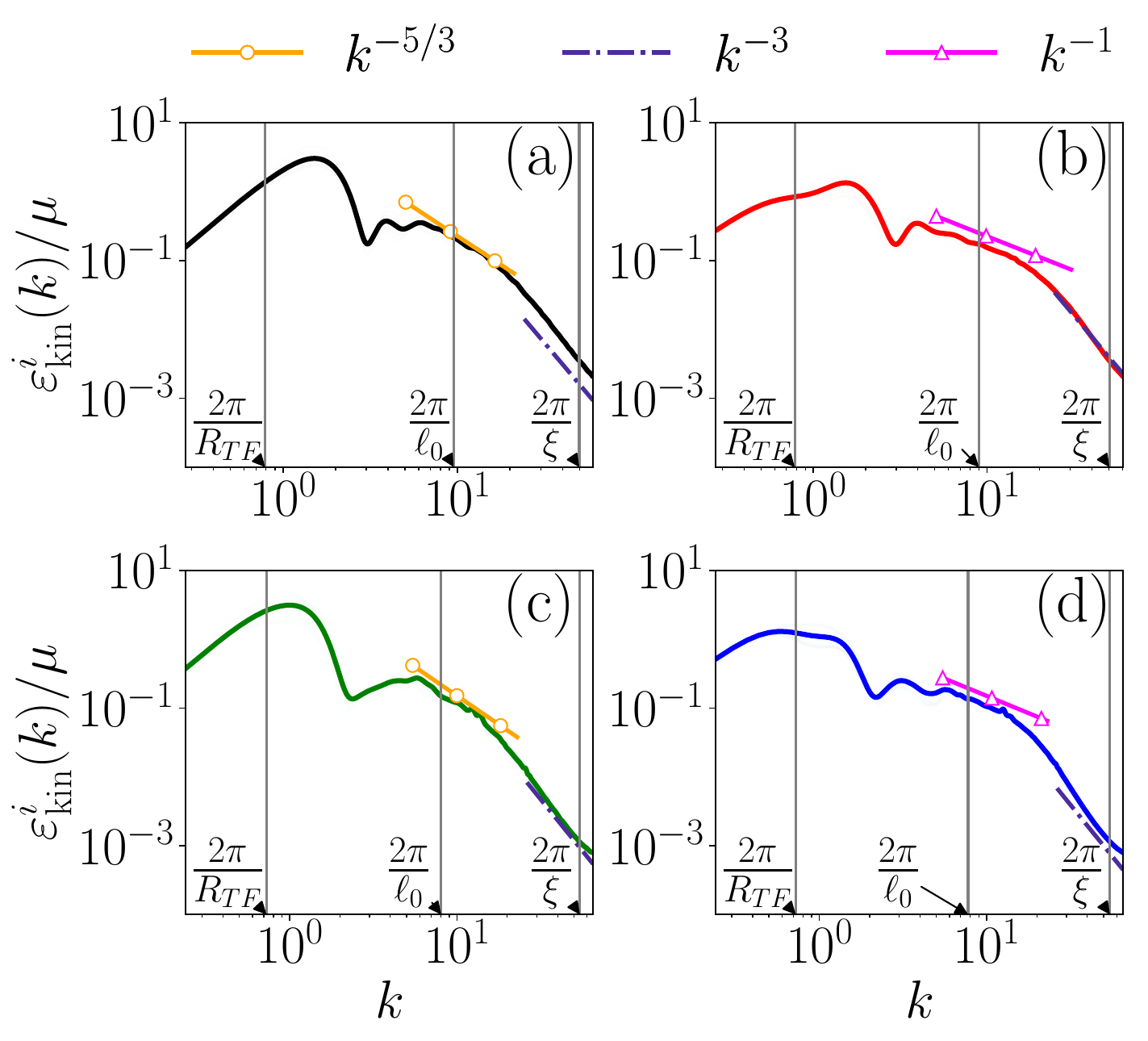}
    \caption{Incompressible kinetic energy spectra for the (a,b) axionic and (c,d) bosonic condensates with $\gdd/\gd = 200$. Spectra are averaged over two time intervals: (a,c) from $t = 2.2$ to $t = 3.2$, and (b,d) from $t = 3.2$ to $t = 5$. Both systems exhibit a Kolmogorov cascade ($k^{-5/3}$) shortly after spin-down, which later transitions to a Vinen turbulence scaling ($k^{-1}$). An enstrophy cascade ($k^{-3}$) is also visible.}
    %\caption{Incompressible kinetic energy spectra for Axionic [(a) and (b)] and bosonic [(c) and (d)] cases with $\gdd/\gd = 200$, averaged from [(a) and (c)] $t = 2.2$ to $t = 3.2$ and [(b) and (d)] from $t=3.2$ to $t=5$. Both cases show Kolmogorov scaling $k^{-5/3}$ after spin-down, which later transitions to Vinen scaling $k^{-1}$. The enstrophy cascade of $k^{-3}$ is also observed.}
    \label{fig:ikinspectra3d}
\end{figure}
%%%%%%%%%%%%%%%%%%%%%%%%%%%%%%%%%%%%%%%%%%%%%%%%%%%%%%%%
It should be noted that our comparison between the bosonic and axionic cases is done for $\gd=\gdd=200$, where the condensate profiles look similar as shown in Fig.~\ref{fig:1dprof}(a). In both cases, the turbulent regime is observed immediately after the spin-down is completed to the final rotation frequency. This is because of the vortices detaching from the pinning sites and joining the surrounding superfluid flow, which reaches a critical velocity threshold. Once vortices enter, they undergo a breakdown into smaller circulations, which leads to the incompressible energy scaling as $k^{-5/3}$ in the inertial range $2\pi/R_{TF} < k < 2\pi/\ell_0$ as shown in the Figs.~\ref{fig:ikinspectra3d}(a) and~\ref{fig:ikinspectra3d}(c). This vortex energy breakdown, fueled by the rotating condensate, is a signature of the Kolmogorov energy cascade. This cascade is prevalent across various turbulent flows ranging from classical fluids to superfluids and BECs~\cite{Tsubota:OUP2025}. Recently, it has been reported in colliding self-gravitating systems~\cite{Sivakumar2025} and even during spin-down of condensates in the presence of a crust-potential~\cite{Sivakumar2026}. Apart from the Kolmogorov cascade, the condensate enstrophy undergoes a self-similar cascade, and manifests itself as a $k^{-3}$ power-law in the large-$k$ ultravoilet regime  in the incompressible spectrum.

The short-lived Kolmogorov regime is succeeded by the Vinen turbulence scaling, the $k^{-1}$ regime at length scales smaller than the intervortex spacing (i.e., $k > 2\pi/\ell_0$) as depicted in  Figs.~\ref{fig:ikinspectra3d}(b) and (d). This transition indicates that the initial vortex population, contributing to the Kolmogorov cascade, decreases, as the $k^{-1}$ scaling is a signature of isolated vortices~\cite{barenghi2023types}. Compared to spin-down turbulence in standard atomic condensates~\cite{Sivakumar2026}, the scaling behavior reported here is relatively weak, because a self-gravitating trap does not retain its shape. So it is challenging to sustain turbulent fluctuations at long length scales (greater than the radius of the collapsed condensate).
%%%%%%%%%%%%%%%%%%%%%%%%%%%%%%%%%%%%%%%%%%%%%%%%%%%%%%%%
\begin{figure}[ht!]
    \centering
    \includegraphics[width=\linewidth]{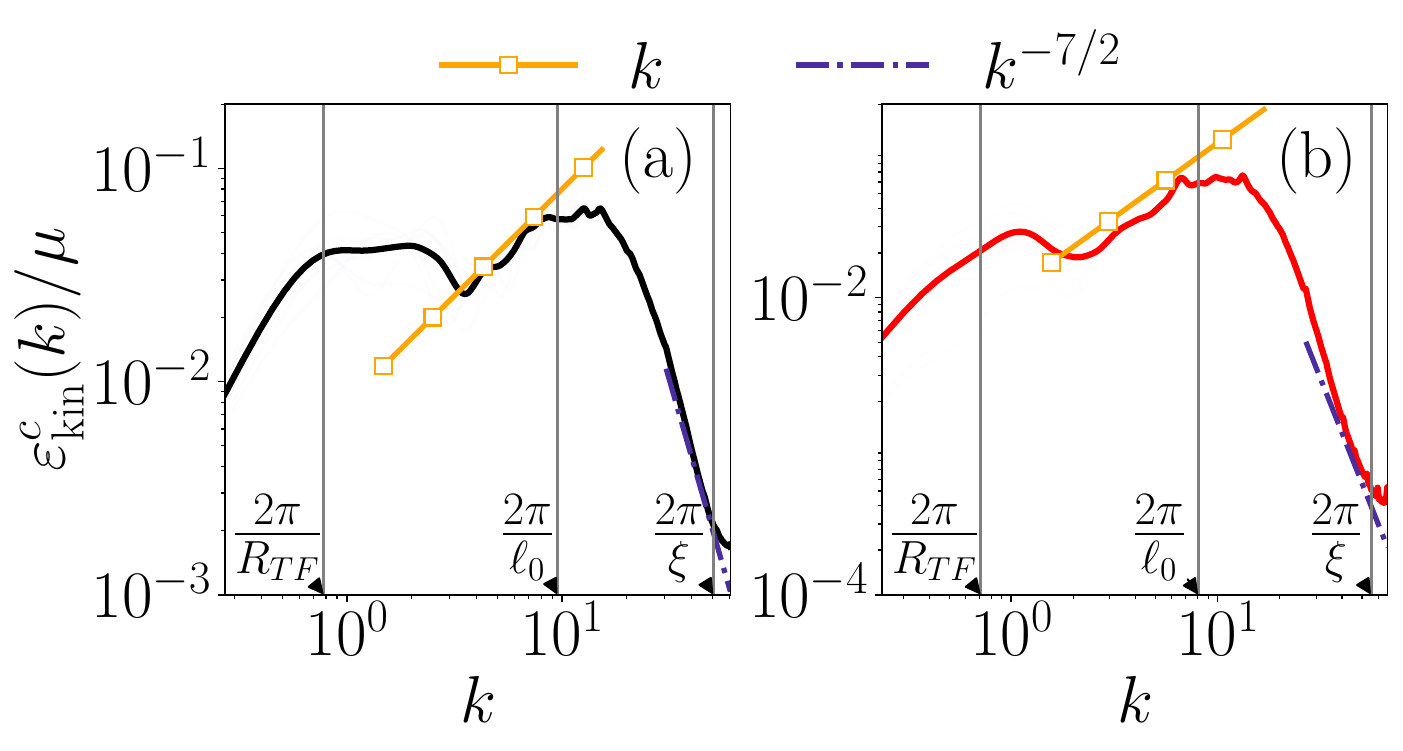}
    \caption{Compressible kinetic energy spectra for (a) the axionic condensate with $\gdd = 200$ and (b) the bosonic condensate with $\gd = 200$, time-averaged over the interval $t=2.2$ to $t=3.2$. Both spectra exhibit a thermalized $k$ scaling within the inertial range [cf. similar spectra in rotating Bose-Einstein condensates~\cite{Estrada2022}].
    %Compressible kinetic energy spectra for (a) Axionic case with $\gdd =200$ and (b) Bosonic case with $\gd = 200$ averaged from averaged from $t=2.2$ to $t=3.2$. Both cases exhibit thermalization with a $k$ scaling in the inertial range.
    }
    \label{fig:ckinspectra}
\end{figure}
%%%%%%%%%%%%%%%%%%%%%%%%%%%%%%%%%%%%%%%%%%%%%%%%%%%%%%%%%%%%%%%%%%

Apart the incompressible kinetic energy spectra, we also analyze the associated compressible energy spectral distribution in Fig.~\ref{fig:ckinspectra}. For the axionic and bosonic cases, it is evident that thermalization is present, indicated by the $k$ scaling  range.  As in the incompressible spectrum, the extent of the $k$ scaling range is reduced by the density-dependent nature of the self-gravitating trap.  Given its larger size, the bosonic condensate displays a longer $k$-scaling range compared to its axionic counterpart. The presence of the crust also helps to sustain the compressible flow around the pinning sites, which makes the $k$ scaling appear in the  range $2\pi/R_{TF} < k < 2\pi/\ell_0$, whereas, in the absence of such a crust, the superfluid flow is expelled to the periphery relatively quickly because of the self-gravitating potential~\cite{Sivakumar2025}. The breakdown of density waves via thermalization is also accompanied by a $k^{-7/2}$ scaling range at large $k$~\cite{AmetteEstrada2022, Sivakumar2024a}.

%%%%%%%%%%%%%%%%%%%%%%%%%%%%%%%%%%%%%%%%%%%%%%%%%%%%%%%%
\begin{figure}[ht!]
    \centering
    \includegraphics[width=\linewidth]{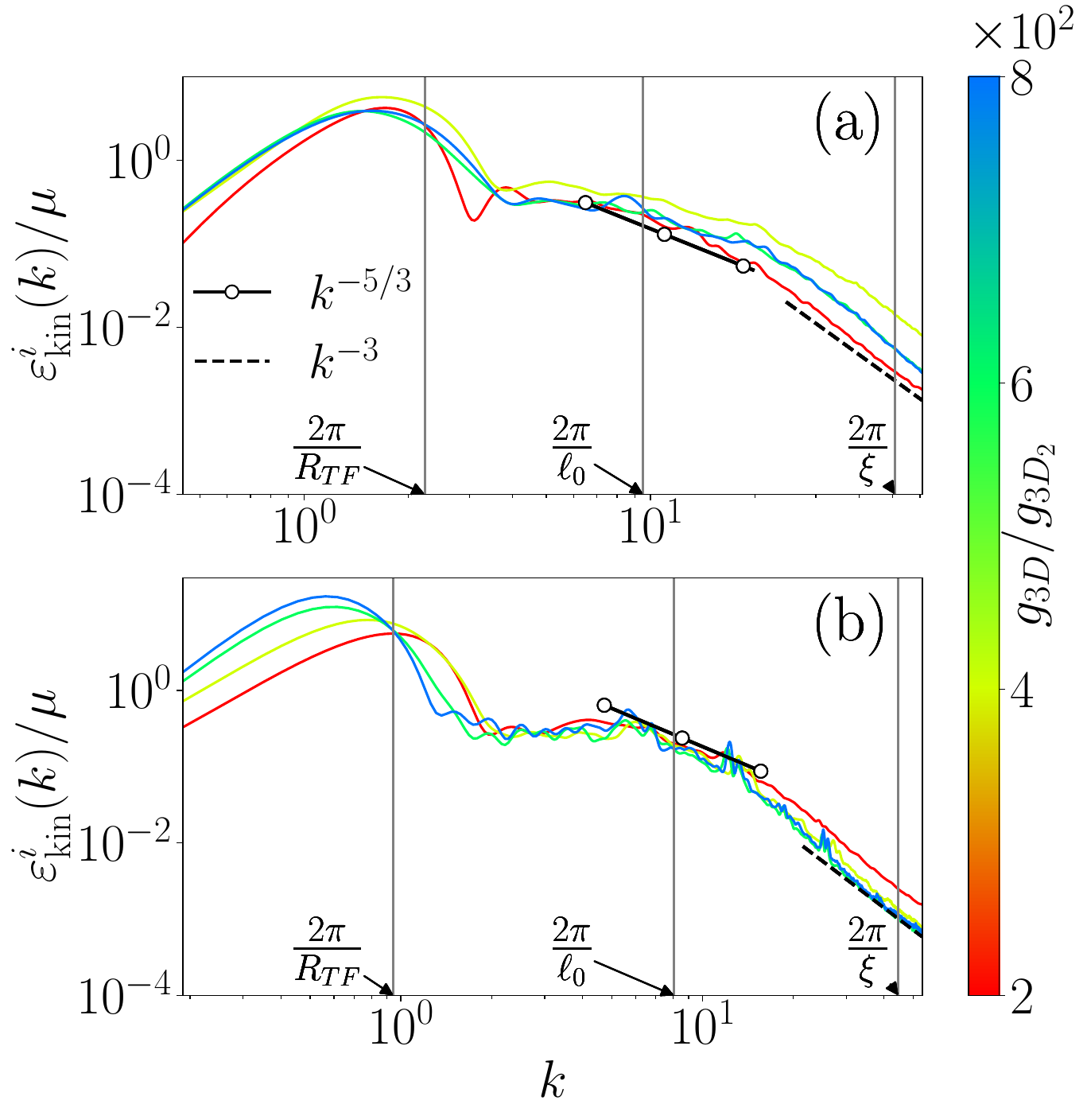}
    \caption{
    Time-averaged incompressible kinetic energy spectra over the interval \( t = 2.2 - 3.2 \) for (a) the axionic case with different \( \gdd \) values and (b) the bosonic case with different \( \gd \) values. The color bar indicates the parameter values.
    %Variation of incompressible kinetic energy spectra averaged over time range $t=2.2$ to $t=3.2$ for (a) Axionic case for different $\gdd$ values (b) Bosonic case for different $\gd$ values.
    }
    \label{fig:spectra_g}
\end{figure}
%%%%%%%%%%%%%%%%%%%%%%%%%%%%%%%%%%%%%%%%%%%%%%%%%%%%%%%%
Although axionic and bosonic cases exhibit similar spectral behaviors for small values of $\gdd$ or $\gd$, there are significant differences as we increase these interaction strengths. For the axionic cases, Fig.~\ref{fig:spectra_g}(a) shows that  there is a deviation from Kolmogorov scaling with increasing $\gdd$. The bosonic case [Fig.~\ref{fig:spectra_g}(b)] also shows Kolmogorov scaling (the smaller the interaction the better the fit to K41); but compared to their axionic counterparts,  the bosonic spectra do not show deviations from K41 if $\gd$ is large. Generally, an increase in the condensate size increases the likelihood of vortices pinning, after immediately depinning; this inhibits the vortex-breakdown process and causes vortex retention. Considering this, the axionic case deviates more from K41 as an increase in $\gdd$ makes it more uniform, thus assisting the vortex retention process. The bosonic case still exhibits the K41 scaling for larger sizes, although it is marked by fluctuations, indicating well-defined vortex structures retained because of the increase in size.

\begin{figure}[ht!]
    \centering
    \includegraphics[width=\linewidth]{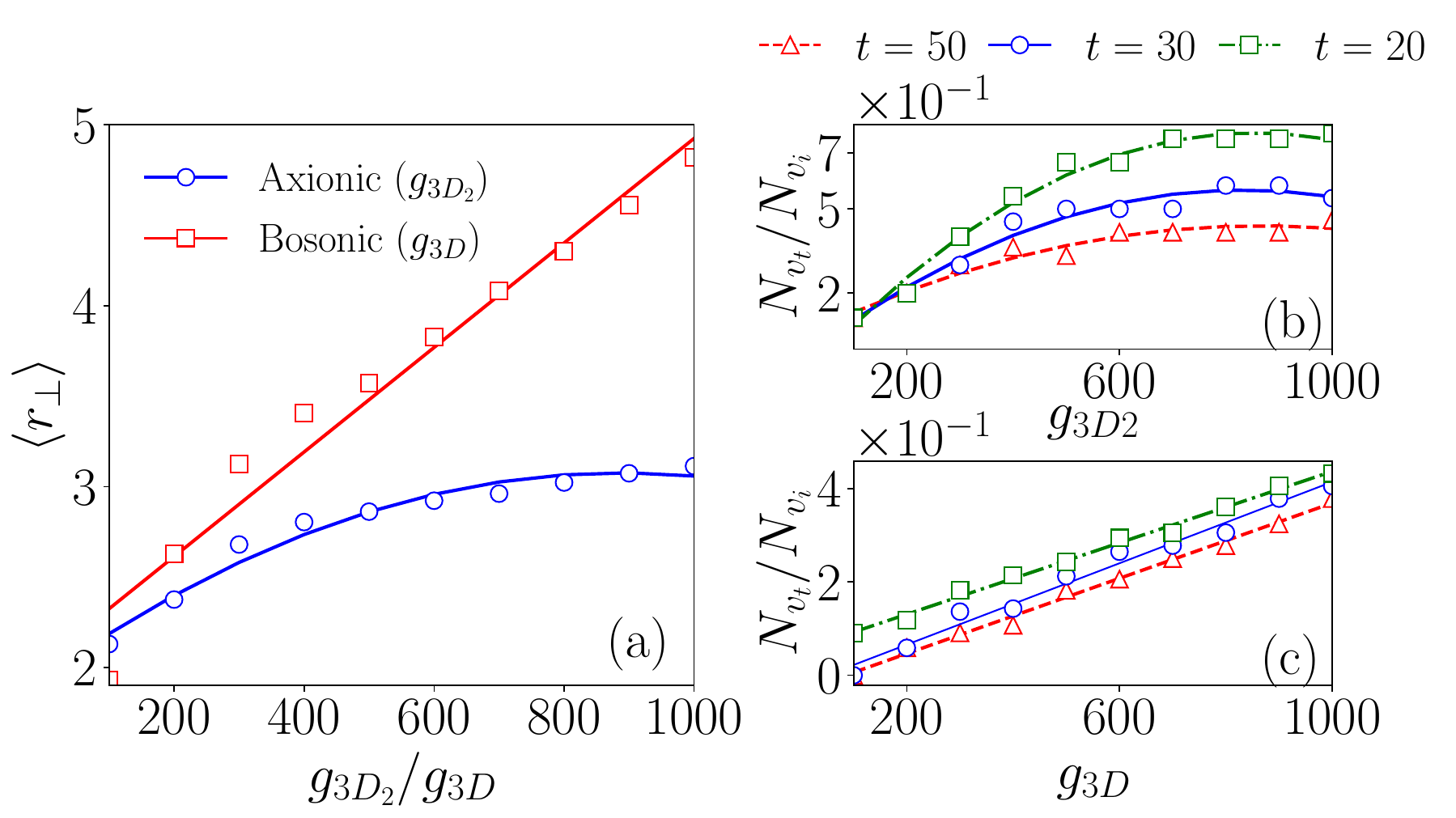}
    \caption{(a) Radial width of the condensates as a function of the interaction strength for the bosonic ($\gd$) and axionic ($\gdd$) cases. Temporal decay of the vortex fraction at times \( t = 20\), \(t=30 \), and \( t = 50 \) for (b) the axionic condensate at different $\gdd$ values and (c) the bosonic condensate at different $\gd$ values.
    %(a) Radial width profile of bosonic and axionic condensates as a function of interaction strength. Time evolution of the vortex fraction at \( t = 20 \), \( t = 30 \), and \( t = 50 \) for (b) the axionic condensate at different \( g_{dd} \) values and (c) the bosonic condensate at different \( g_d\) values.
    }
    \label{fig:vfraction}
\end{figure}

We also observe the average radial width of both types of condensates during the spin-down and note how it varies with their respective interaction strengths in Fig.~\ref{fig:vfraction}(a). The larger radius of the bosonic condensate facilitates more vortices in the condensate, and hence better Kolmogorov scaling compared to the axionic case. Moreover, the radius of the axionic condensate saturates with increasing $\gdd$, hence the difference between axionic and bosonic spectra is also amplified at large interaction strengths.

To corroborate the vortex-retention process, we plot the fraction $N_{v_t}/N_{v_i}$, where $N_{v_t}$ is the vortex population at time $t$ and 
$N_{v_i}$ the initial population, for bosonic and axionic condensates in Figs.~\ref{fig:vfraction}(b) and ~\ref{fig:vfraction}(c), respectively, 
at different interaction strengths. As we have pointed out earlier, the condensate radius increases with the interaction strength, so the latter is an indirect measure of the condensate size. %
%%%%%%%%%%%%%%%%%%%%%%%%%%%%%%%%%%%%%%%%%%%%%%%%%%%%%%%%
%%%%%%%%%%%%%%%%%%%%%%%%%%%%%%%%%%%%%%%%%%%%%%%%%%%%%%
It has been demonstrated in atomic BECs with hard-walled traps that the vortex fraction depends linearly on the initial vortex population, and hence on the condensate size~\cite{Sivakumar2026}. We observe a similar trend here: the vortex fraction varies with the strength of the nonlinearity, much as it does with the condensate radius. For the axionic case, the fraction increases before saturating at large values of $\gdd$ [see Fig.~\ref{fig:vfraction}(b)], whereas it remains linear in the bosonic case [see Fig.~\ref{fig:vfraction}(c)].  The larger a bosonic condensate, the more vortices are retained, because more pinning sites are available; and vortices are more likely to re-pin rather than escape. For the axionic case, the increase may arise from a higher Magnus-flow threshold, which prevents many vortices from depinning at all. The vortex fraction clearly decreases over time as vortices eventually leave the condensate. Even though the axionic condensate is small because its size saturates as $\gdd$ grows, its vortex retention is enhanced, as seen in the significant increase of the vortex fraction [Fig.~\ref{fig:vfraction}(b)].
%It has been previously reported in atomic BECs with hard-walled traps that vortex fraction has a linear relationship with initial vortex population and hence with the condensate size~\cite{Sivakumar2026},so we observe a similar case where the vortex fraction variation with nonlinearity follows the same trend as it does with the condensate radius. The axionic case shows an increase before saturating for larger interaction strengths of $\gdd$ [see Fig.~\ref{fig:vfraction}(b)], while vortex fraction for the bosonic case maintains a linear behaviour [see Fig.~\ref{fig:vfraction}(c)]. More vortices are retained for larger bosonic condensates, as this reveals more pinning sites, which restrict vortex expulsion as they are more likely to re-pin back onto these sites. For axionic cases, this could also be an increase in the Magnus flow threshold, which prevents several vortices from never depinning in the first place. It is also directly evident that the vortex fraction decreases with time as vortices eventually move out of the condensate. Despite the smaller size and its saturation in the axionic case, the vortex retention process seems to be enhanced here, as shown by the significantly higher vortex fraction [Fig.~\ref{fig:vfraction}(b)]. 

\begin{figure}[ht!]
    \centering
    \includegraphics[width=\linewidth]{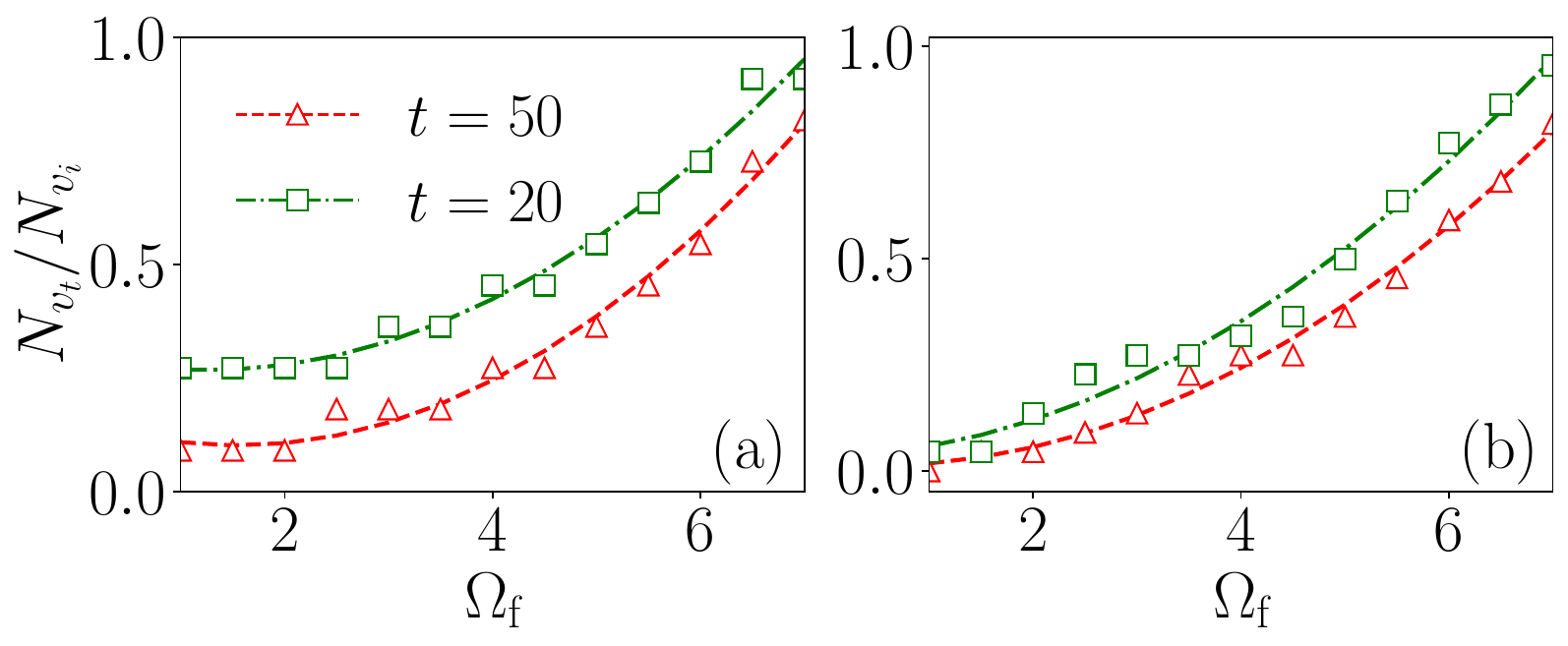}
    \caption{Vortex fraction $N_{v_t}/N_{v_i}$ as a function of the final rotation frequency $\Omega_f$ for the (a) axionic condensate with $\gdd = 200$ and (b) bosonic condensate with $\gd = 200$. }
    %\caption{Variation of vortex fraction with respect to various final rotation frequencies $\Omega_f$ for (a) Axionic case with $\gdd = 200$ and (b) Bosonic case with $\gd = 200$}
    \label{fig:vfraction_spin}
\end{figure}

In order to present a complete picture of vortex retention, we also evaluate the dependence of the vortex-fraction on the final rotation frequency $\Omega_f$  that we depict in Fig.~\ref{fig:vfraction_spin} for the same initial frequency of $\Omega_0$. As we the change $\Omega_f$, the spin-down time also changes because we maintain the same spin-down rate $\dot{\Omega}$, which plays a crucial role in inducing turbulence and vortex explusion~\cite{Liu2025}.
%%%%%%%%%%%%%%%%%%%%%%%%%%%%%%%%%%%%%%%%%%%%
%%%%%%%%%%%%%%%%%%%%%%%%%%%%%%%%%%%%
%This profile is also similar to the case of hard-walled condensates, as it displays an exponential increase in vortex fraction. The interaction strengths of axionic and bosonic cases are taken to be the same lower values of $\gdd = \gd = 200$, which gives us similar profiles same as those observed in Fig.~\ref{fig:vfraction_spin}(a) and \ref{fig:vfraction_spin}(b). The vortex retention process of the axionic case inhibits its expulsion process, which is crucial to obtaining a proper Kolmogorov cascade in our energy spectra. 
This behavior closely resembles that observed in hard-walled condensates, where the vortex fraction exhibits an exponential increase. For a direct comparison, we consider identical and relatively low interaction strengths in both systems, namely, $\gdd=\gd=200$, which yield profiles similar to those shown in in Figs.~\ref{fig:vfraction_spin}(a) and \ref{fig:vfraction_spin}(b). In the axionic case, enhanced vortex retention suppresses vortex expulsion, a process that is essential for sustaining a well-developed Kolmogorov cascade in the energy spectrum.

\begin{figure}[ht!]
    \centering
    \includegraphics[width=\linewidth]{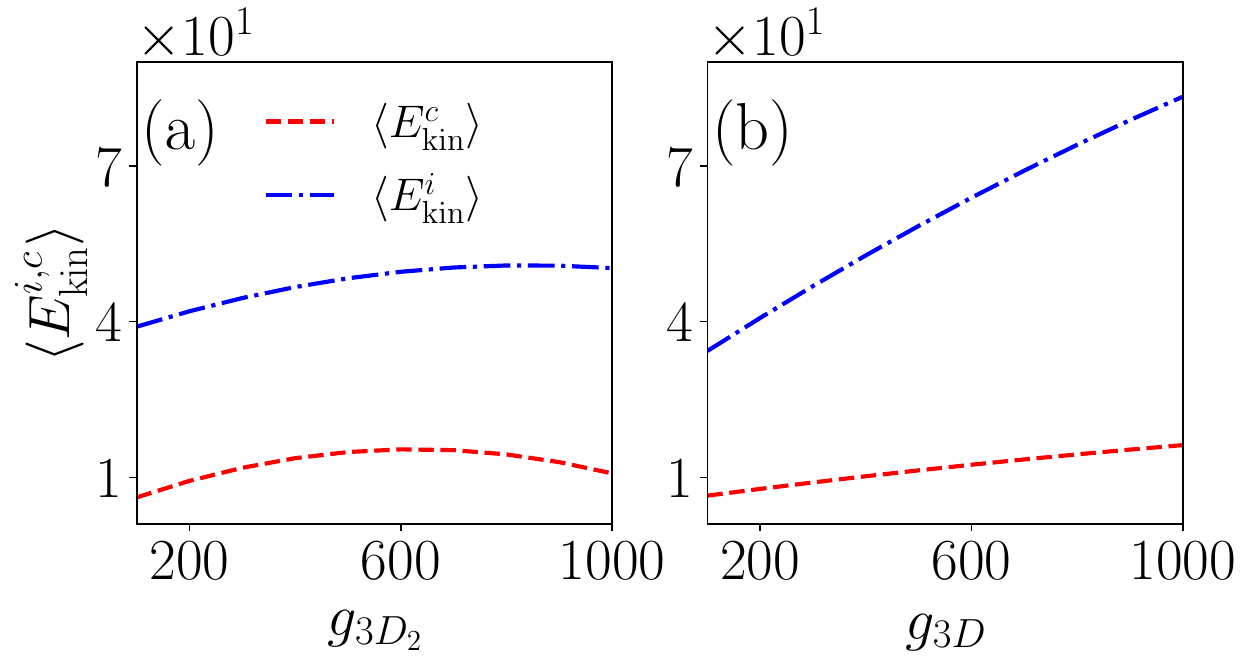}
    \caption{Variation of total incompressible (dash dotted blue line) and compressible (dashed red line) kinetic energies, averaged over $t=t_s$ to $t = 10$ with (a) the axionic interaction strength $\gdd$ and (b) the bosonic interaction strength $\gd$}
    \label{fig:kinener}
\end{figure}

We obtain additional insights into the turbulent dynamics by examining the evolution of the total incompressible and compressible kinetic energy components, $E^{i,c}_\mathrm{kin}$ as a function of the interaction strength. We compute the total kinetic energy within the condensate and average it over the turbulence in the time interval $t_s < t < 10$. As shown in Fig.~\ref{fig:kinener}, the incompressible component dominates over the compressible one across all interaction strengths. Since the condensate radius and vortex fraction serve as indirect measures of the vortex number, and, therefore, of incompressible energy, it is interesting to note that the incompressible energy in the axionic system saturates at large interaction strengths [Fig.~\ref{fig:kinener}(a)], mirroring the behaviors of the radial width [Fig.~\ref{fig:vfraction}(a)] and the vortex fraction [Fig.~\ref{fig:vfraction}(b)]. A similar correspondence between incompressible energy, radial width, and vortex fraction is observed in the bosonic case. Notably, despite its smaller condensate size, the axionic system exhibits a compressible energy comparable to, and in some interaction regimes exceeding, that of the bosonic system. At larger values of $\gdd$, however, the compressible component decreases, indicating enhanced expulsion of compressible excitations by strong axionic interactions.

%Further analysis on the turbulent mechanism can be done by taking a look at the evolution of total incompressible and compressible kinetic energy components ($E^{i,c}_\mathrm{kin} $, with axionic and bosonic interaction strengths. We consider the total kinetic energy component values within the condensate and average this total energy over the turbulent period $t_s < t < 10$. In Fig.~\ref{fig:kinener}, we can observe that the incompressible component dominates over the compressible counterpart for all interaction strengths. Since the condensate radius and vortex fraction are indirect measures for vortex number and hence subsequently the vortex/incompressible energy, it is understandable that the incompressible  component saturates for higher values of axionic interaction [Fig.~\ref{fig:kinener}(a)] similar to radial width [see Fig.~\ref{fig:vfraction}(a)] and vortex fraction [see Fig.~\ref{fig:vfraction}(b)] profiles. Similarly, the incompressible energy for the bosonic case also matches its radial width and vortex fraction profiles. Despite the relatively smaller size, the axionic case matches the compressible distribution of the bosonic counterpart and is even greater for a range of interaction values. For larger $\gdd$, the compressible component decreases, which indicates there is enhanced compressible expulsion for these values.
%%%%%%%%%%%%%%%%%%%%%%%%%%%%%%%%%%%%%%%%%%
\begin{figure}[ht!]
    \centering
    \includegraphics[width=\linewidth]{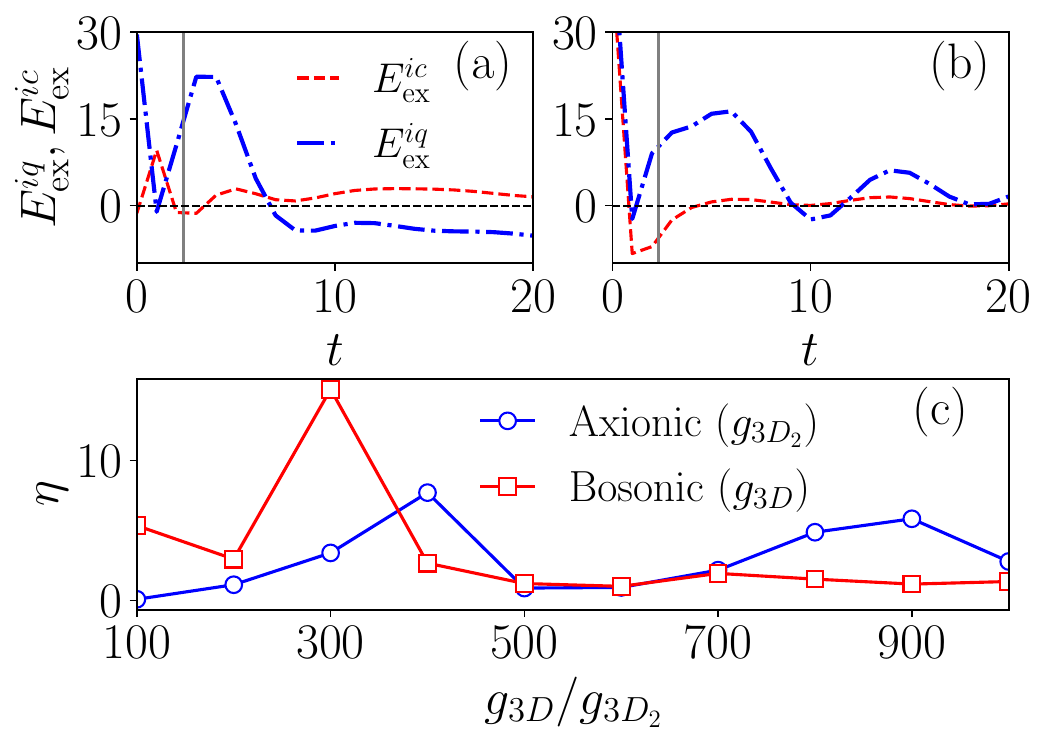}
    \caption{Energy exchange between kinetic energy components. The exchange between incompressible and quantum-pressure components $(E_{\mathrm{kin}}^{iq})$ is shown as a blue dash-dotted line, and between compressible and incompressible components $(E_{\mathrm{kin}}^{ci})$ as a red dashed line. Results are shown for (a) the axionic condensate with $\gdd = 400$ and (b) the bosonic condensate with $\gd = 300$; the solid gray vertical line marks the spin-down time $t_s$. (c) Profile of the energy exchange ratio $\eta$ as a function of the interaction strength for both the bosonic and axionic cases.}
    \label{fig:exchange}
\end{figure}

%%%%%%%%%%%%%%%%%%%%%%%%%%%%%%%%%%%%%%%%%%%%%%%%%
%We further inspect the underlying mechanism of turbulence by studying the exchange between the kinetic energy components. In previous work, where the condensate was completely spun down, an injection of incompressible energy coming from the quantum pressure component was observed~\cite{Sivakumar2026}. The same mechanism is observed here, confirming its association with the crust potential. Even when the condensate is not fully spun down, the injection of energy from quantum pressure remains significant, as shown in the exchange energy plots between the incompressible-quantum pressure components (\(E_\mathrm{kin}^{iq}\)) and the compressible-incompressible components (\(E_\mathrm{kin}^{cq}\)) for both the axionic [Fig.~\ref{fig:exchange}(a)] and bosonic [Fig.~\ref{fig:exchange}(b)] cases. 
Furthermore, we examine the underlying turbulent mechanism by analyzing the exchange between the kinetic energy components. In previous work, where the condensate was fully spun down, an injection of incompressible energy originating from the quantum-pressure term was identified~\cite{Sivakumar2026}. We observe the same mechanism here, confirming its close association with the presence of the crust potential. Even when the condensate is not completely spun down, quantum-pressure-driven energy injection remains significant. This is evident in the energy exchange between the incompressible quantum-pressure components ($E^{iq_{kin}}$) and the compressible-incompressible components ($E^{cq}_{kin}$), as shown in the Fig.~\ref{fig:exchange}(a) for the axionic case and in Fig.~\ref{fig:exchange}(b) for the bosonic case.
%%%%%%%%%%%%%%%%%%%%%%%%%%%%%%%%%%%%%%%%%%%%%%%%%%%%%%%%
%%%%%%%%%%%%%%%%%%%%%%%%%%%%%%%%%%%%%%%%%%%%%%%%%%%%%%
Clearly the growth of the incompressible energy is driven primarily by quantum pressure, rather than by the compressible component, in both systems. %
%%%%%%%%%%%%%%%%%%%%%%%%%%%%%%%%%%%%%%%%%%%%%%%%%
\begin{figure*}[ht!]
    \centering
    \includegraphics[width=\linewidth]{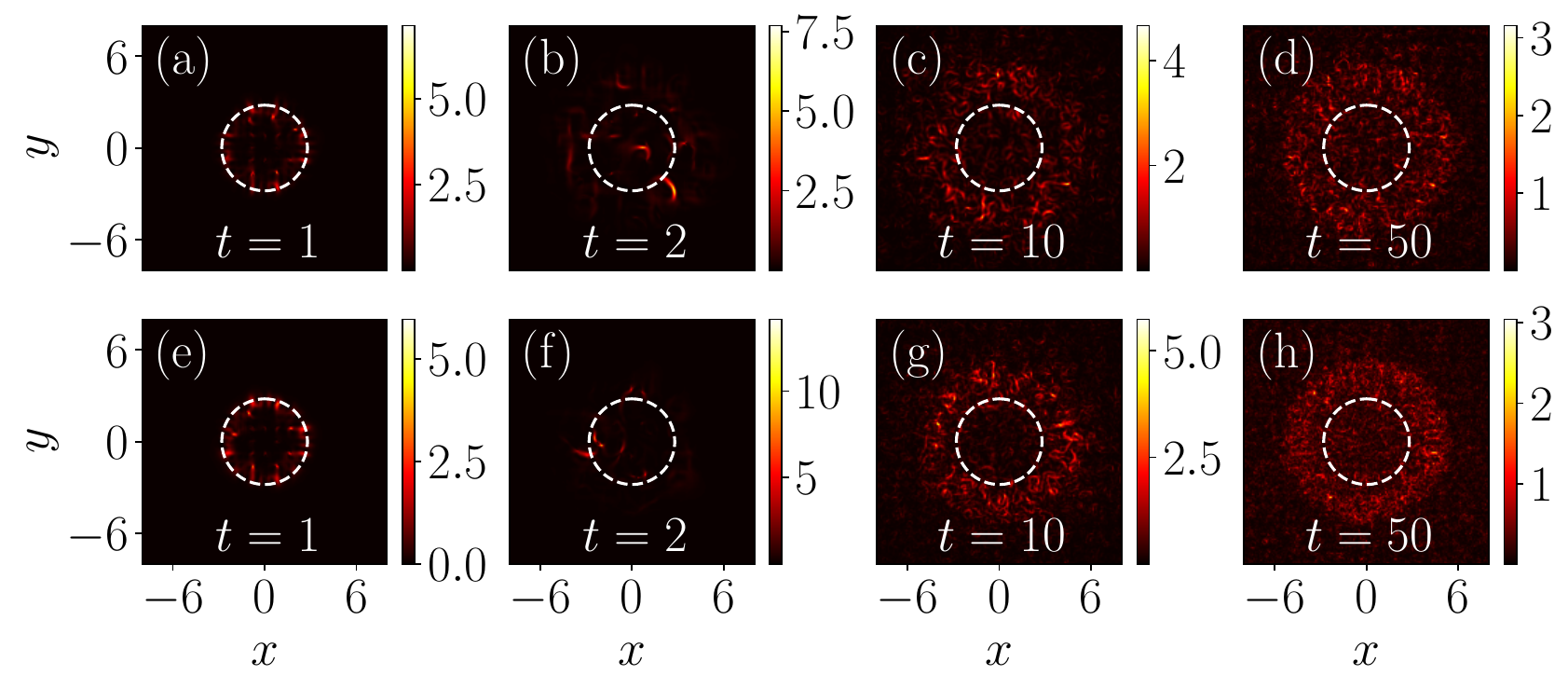}
    \caption{Spatial distribution of the compressible kinetic energy density in the $x$–$y$ midplane ($z=0$). Panels (a)-(d): axionic condensate ($\gdd = 200$) at times $t = 1, 2, 10, 50$. Panels (e)-(h): bosonic condensate ($\gd = 200$) at the same times. The dashed white circle indicates the Thomas-Fermi radius $R\mathrm{TF}$. Energy is initially concentrated at pinning sites and is subsequently expelled to the periphery.}
    \label{fig:spatial}
\end{figure*}%
%%%%%%%%%%%%%%%%%%%%%%%%%%%%%
%This conversion from quantum pressure occurs largely due to the detachment of vortices from their pinning sites, followed by their entry into the surrounding superfluid flow, which manifests as an increase in the incompressible flow component. Previous studies in hard-walled condensates have shown that as condensate size increases, the exchange between compressible and incompressible components becomes more significant relative to the exchange between quantum pressure and the incompressible component~\cite{Sivakumar2026}. This effect stems mainly from the enhanced generation of compressible excitations in larger condensates, which strengthens their coupling to the incompressible energy field. It is also important to note that in hard-walled traps, vortex shedding occurs gradually, and quantum pressure injection can take place before complete spin-down, especially in larger systems. In this self-gravitating system, however, injection consistently occurs after spin-down, indicating simultaneous expulsion even as the condensate size increases.
This conversion from quantum pressure arises principally from vortices detaching from their pinning sites and entering the surrounding superfluid, which manifests itself as an increase in the incompressible flow component. Previous studies in hard-walled condensates have shown that, as the condensate size grows, the exchange between compressible and incompressible components becomes increasingly significant compared to the exchange between quantum pressure and the incompressible component~\cite{Sivakumar2026}. This occurs mainly because of the enhanced generation of compressible excitations in large condensates, which strengthens their coupling to the incompressible energy field. In hard-walled traps, vortex shedding occurs gradually, allowing quantum-pressure injection to begin even before full spin-down, particularly in large systems. In contrast, in bosonic condensates, quantum-pressure-driven injection occurs consistently after spin-down, coinciding with simultaneous vortex and compressible-wave expulsion, even as the condensate size increases.

To quantify the relative dominance between the energy exchange channels, we define the ratio $\eta=(E^{iq}_{ex})_{max}/(E^{ic}_{ex})_{max}$, comparing the peak magnitudes of the quantum--pressure incompressible exchange $E^{iq}_{ex}$ and the compressible-incompressible exchange $E^{ic}_{ex}$ in the the turbulent phase for the time interval $t_s < t < 10$. A large value of $\eta$ indicates that incompressible energy growth is driven more strongly by quantum pressure than by compressible flows. As shown in Fig.~\ref{fig:exchange}(c), $\eta$ exhibits a maximum at an optimal interaction strength for both bosonic and axionic condensates. Beyond this optimal value, increasing the interaction strength and hence the condensate size  enhances the generation of compressible excitations, reducing the relative dominance of $E^{iq}_{ex}$ and lowering $\eta$. Given their compact structure, axionic condensates generally display smaller values of $\eta$ than bosonic condensates of comparable effective size; this reflects their enhanced vortex retention and reduced vortex breakdown.

The spatial distribution of the compressible kinetic energy provides additional insights into the turbulent dynamics of these condensates. As shown in Fig.~\ref{fig:spatial}, compressible excitations originate near the pinning sites [Figs.~\ref{fig:spatial}(a) and \ref{fig:spatial}(e)] and are then advected toward the periphery of the condensate as the system enters the turbulent regime [Figs.~\ref{fig:spatial}(b) and \ref{fig:spatial}(f)]. This outward transport explains the limited range of $k$ scaling in the compressible kinetic energy spectrum: excitations generated near the center are expelled, restricting the scales over which the turbulent cascade develops internally.

At later times, a pronounced halo of compressible energy forms at the condensate edges [Figs.~\ref{fig:spatial}(c) and \ref{fig:spatial}(g)] and is maintained over long durations [Figs.~\ref{fig:spatial}(d) and \ref{fig:spatial}(h)]. This expulsion is a direct consequence of the density-dependent nature of the self-gravitating trap, which lacks confining boundaries to reflect waves back into the core, as previously shown for similar systems~\cite{Sivakumar2025}. Consequently, the compressible energy density within the central condensate region becomes minimal at later times, with the remaining signal primarily arising from superfluid flow perturbations around the residual pinning sites.

The morphology of the compressible energy also reflects the role of rotation. Without rotation, compressible excitations fragment into small, disordered structures. In contrast, rotation organizes them into coherent, strand-like clumps~\cite{Sivakumar2024a}. Consistently, we observe that, after the initial turbulent phase, compressible energy coalesces into strands [Figs.~\ref{fig:spatial}(c) and \ref{fig:spatial}(g)], which gradually break down into finer structures as the rotation frequency continues to decrease during spin-down.

\section{Summary and Conclusions}
\label{sec:summary}
We have conducted a numerical study of turbulent spin-down in self-gravitating Bose Einstein condensates, comparing standard bosonic systems with axionic condensates featuring higher-order interactions. Gross-Pitaevskii-Poisson simulations with a pinning crust potential show that axionic interactions produce a more uniform condensate and lower the critical rotation frequency for vortex entry. Rapid spin-down induces a Magnus flow which, upon exceeding a critical velocity, triggers large-scale vortex depinning and seeds turbulence. While vortex avalanches in pinned systems are commonly invoked in neutron-star analogs, self-gravitating condensates without pinning can also expel vortices once the circulation exceeds a critical threshold \(\xi_c\)~\cite{Nikolaieva2023, Sivakumar2025}. These behaviors can be unified through a circulation-dependent Magnus-flow criterion~\cite{Liu2025, Sonin1997}. Turbulence is characterized using compressible and incompressible kinetic energy spectra. For direct comparison, we set \(\gdd = \gd = 200\), yielding similar density profiles in both bosonic and axionic systems.

For condensates tuned to similar effective sizes, both bosonic and axionic systems follow a common turbulent pathway: a short inertial range with a Kolmogorov-like $k^{-5/3}$ cascade that makes a transition to the Vinen-like scaling $k^{-1}$ as the vortex tangle decays. The compressible spectrum consistently exhibits signatures of thermalization with scalings ($\sim k$) and ($\sim k^{-7/2}$). For this case, the scaling range is limited compared to that in atomic BECs because of vortex expulsion and density fluctuations arising from the density-dependent trapping potential.

Differences emerge at high interaction strengths, where the increasing condensate size enhances vortex retention in both systems, but much more strongly in the axionic case. This leads to a pronounced suppression and deviation from Kolmogorov scaling in the axionic energy spectrum at large \(\gdd\), while the bosonic spectrum largely retains the cascade with increased fluctuations. This behavior correlates with the condensate structure: although the radial width grows linearly with \(\gd\) in both cases, axionic condensates grow more slowly and eventually saturate. Despite their smaller size, axionic systems retain a higher vortex fraction because of the delayed depinning from elevated Magnus-flow thresholds. Enhanced vortex retention suppresses vortex expulsion and breakdown, thereby inhibiting the classical Kolmogorov cascade at high interaction strengths.

To elucidate the underlying turbulent mechanism, we have analyzed the evolution of incompressible and compressible kinetic energies during the turbulent and post-turbulent phase ($t_s < t < 10$) as functions of the interaction strength. The total incompressible energy closely tracks both the vortex fraction and the condensate radius, reflecting their direct dependence on the vortex population. In axionic systems, however, the total compressible energy decreases at large interaction strengths, indicating enhanced expulsion of density fluctuations and suggesting an optimal axionic regime for retaining compressible excitations. 

Energy-exchange analysis shows that the growth of the incompressible flow is primarily driven by quantum pressure injection in both systems, originating from vortex depinning and entry into the superfluid. This effect is stronger in axionic condensates given their sharper density gradients at vortex cores. Unlike hard-walled traps, where compressible-to-incompressible exchange can dominate in large condensates~\cite{Sivakumar2026}, self-gravitating systems suppress this channel because the density-dependent trap lacks reflective boundaries, preventing sustained compressible-wave amplification.

To study the relative dominance between these energy exchange channels, we have used the parameter $\eta = (E_\mathrm{ex}^{iq})\mathrm{max} / (E_\mathrm{ex}^{ic})_\mathrm{max}$, which quantifies the relative dominance of quantum-pressure-driven versus compressible-driven incompressible growth. While bosonic systems exhibit a higher peak in $\eta$, axionic systems show reduced values despite stronger quantum-pressure injection, because of the simultaneous enhancement of compressible excitations. At higher values of $\gdd$ compressible-wave expulsion produces a secondary peak in $\eta$, consistent with the decline in the total compressible energy.

Overall, we have demonstrated that both bosonic and axionic self-gravitating BECs display universal turbulent features: vortex depinning, a transient quantum-pressure-driven Kolmogorov cascade, and decay into Vinen turbulence. However, higher-order axionic interactions promote a denser, more uniform core, which enhances vortex retention and suppresses expulsion, thereby weakening the forward cascade relative to its counterpart in the bosonic case. These results highlight the critical role of the microscopic interaction potentials  in shaping the large-scale turbulent dissipation in astrophysical superfluids.

\acknowledgments
A.S. acknowledges financial support from the Council of Scientific and Industrial Research (CSIR), India, in the form of a Direct Senior Research Fellowship. The work of P.M. is supported by the Ministry of Education-Rashtriya Uchchatar Shiksha Abhiyan (MoE RUSA 2.0): Bharathidasan University -- Physical Sciences. R.P. and S.S. thank the Anusandhan National Research Foundation (ANRF), India for support.

\appendix

\section{Numerical Solution of the Gross--Pitaevskii--Poisson System Using a Combined Split-Step Crank--Nicolson and Fourier Transform Method}
\label{sec:appendix:a}

In this appendix, we describe the numerical scheme used to solve the coupled Gross--Pitaevskii--Poisson system, Eqs.~\eqref{eq:gpe} and \eqref{eq:poisson}. At each iteration, the Poisson equation~\eqref{eq:poisson} is first solved using the Poisson solver described in Sec.~\ref{app:poisson} to obtain the gravitational potential. The resulting potential is then used in the time integration of the GP equation~\eqref{eq:gpe}, which is performed using the split-step Crank--Nicolson method as detailed in Sec.~\ref{app:sscn}.

\subsection{Split-Step Crank--Nicolson method}
\label{app:sscn}
For the split-step integration, the GP equation is written in operator-split form as
\begin{align}\label{eq:gpe:sscn_main}
\mathrm{i}\,\frac{\partial \psi}{\partial t}
= \left( H_1 + H_2 + H_3 + H_4 \right)\psi \equiv H\psi,
\end{align}
where
\begin{subequations}\label{eq:gpe:sscn_ops}
\begin{align}
H_1 &= V_\mathrm{conf} + V_\mathrm{int}, \label{eq:gpe:sscn:a} \\
H_2 &= -\frac{1}{2}\frac{\partial^2}{\partial x^2}
      - \mathrm{i}\,\Omega(t)\,y\,\frac{\partial}{\partial x}, \label{eq:gpe:sscn:b} \\
H_3 &= -\frac{1}{2}\frac{\partial^2}{\partial y^2}
      + \mathrm{i}\,\Omega(t)\,x\,\frac{\partial}{\partial y}, \label{eq:gpe:sscn:c} \\
H_4 &= -\frac{1}{2}\frac{\partial^2}{\partial z^2}. \label{eq:gpe:sscn:d}
\end{align}
\end{subequations}
Here, \(H_1\) contains the potential and interaction terms, while \(H_2\), \(H_3\), and \(H_4\) represent the directional kinetic and rotational operators used in the operator-splitting scheme.

The time evolution over a small time step $\Delta t$ is approximated using a second-order Strang splitting of the full propagator,
\begin{align}
\psi(\mathbf r, t+\Delta t)
\approx e^{-i H \Delta t}\psi(\mathbf r,t),
\end{align}
where the Hamiltonian is decomposed into potential, kinetic, and rotational contributions. The time propagation is implemented as a sequence of exponential operators acting on the wavefunction, following the standard split-step Crank-Nicolson framework described in Refs.~\cite{Muruganandam2009, Vudragovic2012, Loncar2016, YoungS2017, Kumar2019, Muruganandam2025}.

The first stage corresponds to the non-derivative (local) contribution,
\begin{align}
\mathrm i\frac{\partial \psi}{\partial t}=H_1\psi,
\end{align}
which is solved exactly over the time step to give
\begin{align}
\psi^{(1)}(\mathbf r)
=
\mathrm e^{-\mathrm i H_1 \Delta t}\psi(\mathbf r,t_n).
\end{align}
Since $H_1$ contains no spatial derivatives, this step is evaluated pointwise in real space.

Second, the derivative step associated with the operator $H_2$ is applied to the intermediate solution using the split-step Crank-Nicolson discretization. The evolution equation
\begin{align}
i\,\frac{\partial \psi}{\partial t} = H_2 \psi
\end{align}
is advanced over one time step $\Delta t$ using the semi-implicit Crank--Nicolson scheme,
\begin{align}
\frac{\psi^{n+1} - \psi^{n}}{-\mathrm i\,\Delta t} = \frac{1}{2} H_2 \left(\psi^{n+1} + \psi^{n}\right),
\end{align}
which yields the update
\begin{align}
\psi^{n+1} = \frac{1 - \mathrm i\,\Delta t\, H_2 / 2}{1 + \mathrm i\,\Delta t\, H_2 / 2}\,
\psi^{n}.
\end{align}
To illustrate the discretization, consider the action of $H_2$ along the $x$ direction. The spatial domain is discretized uniformly over $[x_{\min}, x_{\max}]$ with grid points
\begin{align}
x_i = x_{\min} + i\,h_x, \quad i=0,1,\ldots,N_x,
\end{align}
where $h_x=(x_{\max}-x_{\min})/N_x$ is the spatial step size. The first spatial derivative appearing in the rotational term of $H_2$ is
discretized using a second-order central finite-difference approximation.
At the grid point $x_i$, it is written as
\begin{align}
\left.\frac{\partial \psi}{\partial x}\right|_{x_i}
\approx
\frac{\psi_{i+1}-\psi_{i-1}}{2h_x},
\end{align}
where $h_x$ is the spatial grid spacing and $\psi_i=\psi(x_i)$.

Similarly, the second-order spatial derivative appearing in $H_2$ is approximated using the three-point central finite-difference formul,
\begin{align}
\left.\frac{\partial^2 \psi}{\partial x^2}\right|_{x_i}
\approx
\frac{\psi_{i+1}-2\psi_i+\psi_{i-1}}{h_x^2}.
\end{align}
Substitution into the Crank-Nicolson time discretization yields a tridiagonal system of linear algebraic equations for the updated wavefunction along the $x$ direction. This system is solved efficiently using a forward-backward recursion (Thomas) algorithm, with computational cost scaling linearly with the number of grid points \cite{Koonin2018}. Boundary conditions are imposed such that the wavefunction vanishes at the edges of the computational domain.

After completing the $x$-direction update, the same procedure is applied sequentially to the operators $H_3$ and $H_4$ along the $y$ and $z$ directions, respectively. This completes one full time step of the split-step Crank-Nicolson evolution.

At each iteration, the Poisson equation~\eqref{eq:poisson} is solved to compute the potential using the Poisson solver described below. The resulting potential is then employed in the numerical solution of the GP equation~\eqref{eq:gpe} via the split-step Crank–Nicolson method described above.

\subsection{Implementation of Poisson solver}
\label{app:poisson}
The gravitational potential is computed by solving the Poisson equation~\eqref{eq:poisson} using a spectral collocation approach.
We define the residual $R \equiv R(\mathbf r,t)$ as
\begin{align}
R(\mathbf r,t)=\nabla^2 \Phi(\mathbf r,t)-|\psi(\mathbf r,t)|^2,
\end{align}
where $\Phi$ denotes the numerical approximation to the gravitational
potential.

The approximate solution is expanded in terms of trial basis functions $u_{ijk}(\mathbf r)$ as
\begin{align}
\Phi(\mathbf r,t)
=
\sum_{i=0}^{N_r-1}
\sum_{j=0}^{N_r-1}
\sum_{k=0}^{N_r-1}
\hat{\phi}_{ijk}(t)\,u_{ijk}(\mathbf r),
\end{align}
where $N_r$ is the number of grid points along each spatial direction.
The expansion coefficients $\hat{\phi}_{ijk}$ are determined by enforcing the residual to vanish at the collocation points $\mathbf r_{lmn}=(x_l,y_m,z_n)$.
This corresponds to choosing test functions
\begin{align}
\chi_{lmn}(\mathbf r)=\delta(\mathbf r-\mathbf r_{lmn}),
\end{align}
with $\delta$ denoting the Dirac delta function.

Applying the collocation condition $R(\mathbf r_{lmn},t)=0$ yields
\begin{align}
\sum_{i,j,k=0}^{N_r-1}
\hat{\phi}_{ijk}\,
\nabla^2 u_{ijk}(\mathbf r_{lmn})
-
|\psi(\mathbf r_{lmn})|^2
=0.
\label{eq:app_residue}
\end{align}

To proceed, we choose Fourier basis functions of the form
\begin{align}
u_{ijk}(\mathbf r)
=
\exp\!\left(-\mathrm i\,\mathbf k_{ijk}\cdot\mathbf r\right),
\end{align}
where $\mathbf k_{ijk}=(k_{x_i},k_{y_j},k_{z_k})$ are discrete
three-dimensional wave vectors.
Assuming periodic boundary conditions, the wave numbers are defined as
\begin{align}
k_{x_i}=\frac{2\pi i}{L_x},\quad
k_{y_j}=\frac{2\pi j}{L_y},\quad
k_{z_k}=\frac{2\pi k}{L_z},
\end{align}
with $L_\alpha=N_r h$ $(\alpha=x,y,z)$ and
$i,j,k=-\frac{N_r}{2},\ldots,\frac{N_r}{2}-1$.

The three-dimensional discrete Fourier transform (DFT) and its inverse are
defined by
\begin{align}
\hat{\phi}_{ijk}
&=
\frac{1}{N_r^{3/2}}
\sum_{l,m,n=0}^{N_r-1}
\phi_{lmn}\,
\exp\!\left[
2\pi \mathrm i\frac{il+jm+kn}{N_r}
\right],
\end{align}
\begin{align}
\phi_{lmn}
&=
\frac{1}{N_r^{3/2}}
\sum_{i,j,k=-N_r/2}^{N_r/2-1}
\hat{\phi}_{ijk}\,
\exp\!\left[
-2\pi \mathrm i\frac{il+jm+kn}{N_r}
\right],
\end{align}
where $\phi_{lmn}=\Phi(\mathbf r_{lmn})$.
These transforms satisfy the discrete orthogonality relation
\begin{align}
\frac{1}{N_r^3}
\sum_{l,m,n=0}^{N_r-1}
\exp\!\left[ 2\pi \mathrm i\frac{(i-i')l+(j-j')m+(k-k')n}{N_r} \right] \notag \\
= \delta_{ii'}\delta_{jj'}\delta_{kk'},
\end{align}
ensuring that
\begin{align}
\phi_{lmn}
=
\mathcal F^{-1}\!\left[\mathcal F(\phi_{lmn})\right].
\end{align}

Since Fourier modes are eigenfunctions of the Laplacian operator,
\begin{align}
\nabla^2
\exp(-\mathrm i\,\mathbf k_{ijk}\cdot\mathbf r)
=
-|\mathbf k_{ijk}|^2
\exp(-\mathrm i\,\mathbf k_{ijk}\cdot\mathbf r),
\end{align}
where
\begin{align}
|\mathbf k_{ijk}|^2
=
k_{x_i}^2+k_{y_j}^2+k_{z_k}^2.
\end{align}

Applying the inverse DFT to the first term of
Eq.~\eqref{eq:app_residue} gives
\begin{align}
\mathcal F^{-1}
\!\left[
-|\mathbf k_{ijk}|^2\,\hat{\phi}_{ijk}
\right]
-
|\psi(\mathbf r_{lmn})|^2
=0.
\end{align}
Taking the DFT of this equation yields the spectral Poisson equation
\begin{align}
-|\mathbf k_{ijk}|^2\,\hat{\phi}_{ijk}
= \mathcal F\!\left[|\psi(\mathbf r_{lmn})|^2\right].
\end{align}
For nonzero wave numbers, the Fourier coefficients of the gravitational
potential are therefore obtained as
\begin{align}
\hat{\phi}_{ijk} = - \frac{\mathcal F\!\left[|\psi(\mathbf r_{lmn})|^2\right]}{|\mathbf k_{ijk}|^2}, \qquad \mathbf k_{ijk} \neq 0.
\end{align}
The zero mode is fixed by imposing a suitable reference condition on the potential.

Finally, the gravitational potential in real space is recovered by the inverse transform,
\begin{align}
\Phi(\mathbf r_{lmn}) = \mathcal F^{-1} \!\left[ - \frac{\mathcal F(|\psi|^2)}{|\mathbf k_{ijk}|^2}
\right].
\end{align}

Both the forward and inverse discrete Fourier transforms are implemented using the Fast Fourier Transform (FFT) algorithm. In the present work, GPU-accelerated FFTs are performed using the \texttt{cuFFT} library.

The above numerical procedures were implemented using CUDA C programming \cite{Loncar2016} on an NVIDIA A100 GPU. The simulations were carried out on three-dimensional grids of suitable resolution (e.g., $256\times 256\times 256$) with appropriate spatial and temporal step sizes.


\begin{thebibliography}{77}%
\makeatletter
\providecommand \@ifxundefined [1]{%
 \@ifx{#1\undefined}
}%
\providecommand \@ifnum [1]{%
 \ifnum #1\expandafter \@firstoftwo
 \else \expandafter \@secondoftwo
 \fi
}%
\providecommand \@ifx [1]{%
 \ifx #1\expandafter \@firstoftwo
 \else \expandafter \@secondoftwo
 \fi
}%
\providecommand \natexlab [1]{#1}%
\providecommand \enquote  [1]{``#1''}%
\providecommand \bibnamefont  [1]{#1}%
\providecommand \bibfnamefont [1]{#1}%
\providecommand \citenamefont [1]{#1}%
\providecommand \href@noop [0]{\@secondoftwo}%
\providecommand \href [0]{\begingroup \@sanitize@url \@href}%
\providecommand \@href[1]{\@@startlink{#1}\@@href}%
\providecommand \@@href[1]{\endgroup#1\@@endlink}%
\providecommand \@sanitize@url [0]{\catcode `\\12\catcode `\$12\catcode
  `\&12\catcode `\#12\catcode `\^12\catcode `\_12\catcode `\%12\relax}%
\providecommand \@@startlink[1]{}%
\providecommand \@@endlink[0]{}%
\providecommand \url  [0]{\begingroup\@sanitize@url \@url }%
\providecommand \@url [1]{\endgroup\@href {#1}{\urlprefix }}%
\providecommand \urlprefix  [0]{URL }%
\providecommand \Eprint [0]{\href }%
\providecommand \doibase [0]{https://doi.org/}%
\providecommand \selectlanguage [0]{\@gobble}%
\providecommand \bibinfo  [0]{\@secondoftwo}%
\providecommand \bibfield  [0]{\@secondoftwo}%
\providecommand \translation [1]{[#1]}%
\providecommand \BibitemOpen [0]{}%
\providecommand \bibitemStop [0]{}%
\providecommand \bibitemNoStop [0]{.\EOS\space}%
\providecommand \EOS [0]{\spacefactor3000\relax}%
\providecommand \BibitemShut  [1]{\csname bibitem#1\endcsname}%
\let\auto@bib@innerbib\@empty
%</preamble>
\bibitem [{\citenamefont {Feynman}(1955)}]{Feynman1955}%
  \BibitemOpen
  \bibfield  {author} {\bibinfo {author} {\bibfnamefont {R.~P.}\ \bibnamefont
  {Feynman}},\ }\href {https://doi.org/10.1016/s0079-6417(08)60077-3} {\bibinfo
  {title} {{Chapter II: Application} of quantum mechanics to liquid helium}}
  (\bibinfo {year} {1955})\BibitemShut {NoStop}%
\bibitem [{\citenamefont {Vinen}(1957)}]{Vinen1957}%
  \BibitemOpen
  \bibfield  {author} {\bibinfo {author} {\bibfnamefont {W.~F.}\ \bibnamefont
  {Vinen}},\ }\bibfield  {title} {\bibinfo {title} {Mutual friction in a heat
  current in liquid helium ii i. experiments on steady heat currents},\ }\href
  {https://doi.org/10.1098/rspa.1957.0071} {\bibfield  {journal} {\bibinfo
  {journal} {Proc. R. Soc. A.}\ }\textbf {\bibinfo {volume} {240}},\ \bibinfo
  {pages} {114} (\bibinfo {year} {1957})}\BibitemShut {NoStop}%
\bibitem [{\citenamefont {Paoletti}\ and\ \citenamefont
  {Lathrop}(2011)}]{paoletti2011quantum}%
  \BibitemOpen
  \bibfield  {author} {\bibinfo {author} {\bibfnamefont {M.~S.}\ \bibnamefont
  {Paoletti}}\ and\ \bibinfo {author} {\bibfnamefont {D.~P.}\ \bibnamefont
  {Lathrop}},\ }\bibfield  {title} {\bibinfo {title} {Quantum turbulence},\
  }\href@noop {} {\bibfield  {journal} {\bibinfo  {journal} {Annu. Rev.
  Condens. Matter Phys.}\ }\textbf {\bibinfo {volume} {2}},\ \bibinfo {pages}
  {213} (\bibinfo {year} {2011})}\BibitemShut {NoStop}%
\bibitem [{\citenamefont {Barenghi}\ \emph
  {et~al.}(2023{\natexlab{a}})\citenamefont {Barenghi}, \citenamefont
  {Skrbek},\ and\ \citenamefont {Sreenivasan}}]{Barenghi:CUP2023}%
  \BibitemOpen
  \bibfield  {author} {\bibinfo {author} {\bibfnamefont {C.~F.}\ \bibnamefont
  {Barenghi}}, \bibinfo {author} {\bibfnamefont {L.}~\bibnamefont {Skrbek}},\
  and\ \bibinfo {author} {\bibfnamefont {K.~R.}\ \bibnamefont {Sreenivasan}},\
  }\href {https://doi.org/10.1017/9781009345651} {\emph {\bibinfo {title}
  {Quantum Turbulence}}}\ (\bibinfo  {publisher} {Cambridge University Press},\
  \bibinfo {year} {2023})\BibitemShut {NoStop}%
\bibitem [{\citenamefont {Tsubota}\ and\ \citenamefont
  {Kasamatsu}(2025)}]{Tsubota:OUP2025}%
  \BibitemOpen
  \bibfield  {author} {\bibinfo {author} {\bibfnamefont {M.}~\bibnamefont
  {Tsubota}}\ and\ \bibinfo {author} {\bibfnamefont {K.}~\bibnamefont
  {Kasamatsu}},\ }\href@noop {} {\emph {\bibinfo {title} {Quantum Hydrodynamics
  and Turbulence}}}\ (\bibinfo  {publisher} {Oxford University Press},\
  \bibinfo {year} {2025})\ p.\ \bibinfo {pages} {384}\BibitemShut {NoStop}%
\bibitem [{\citenamefont {White}\ \emph {et~al.}(2014)\citenamefont {White},
  \citenamefont {Anderson},\ and\ \citenamefont {Bagnato}}]{White2014}%
  \BibitemOpen
  \bibfield  {author} {\bibinfo {author} {\bibfnamefont {A.~C.}\ \bibnamefont
  {White}}, \bibinfo {author} {\bibfnamefont {B.~P.}\ \bibnamefont
  {Anderson}},\ and\ \bibinfo {author} {\bibfnamefont {V.~S.}\ \bibnamefont
  {Bagnato}},\ }\bibfield  {title} {\bibinfo {title} {Vortices and turbulence
  in trapped atomic condensates},\ }\href
  {https://doi.org/10.1073/pnas.1312737110} {\bibfield  {journal} {\bibinfo
  {journal} {Proc. Natl. Acad. Sci.}\ }\textbf {\bibinfo {volume} {111}},\
  \bibinfo {pages} {4719} (\bibinfo {year} {2014})}\BibitemShut {NoStop}%
\bibitem [{\citenamefont {Guo}\ \emph {et~al.}(2010)\citenamefont {Guo},
  \citenamefont {Cahn}, \citenamefont {Nikkel}, \citenamefont {Vinen},\ and\
  \citenamefont {McKinsey}}]{Guo2010}%
  \BibitemOpen
  \bibfield  {author} {\bibinfo {author} {\bibfnamefont {W.}~\bibnamefont
  {Guo}}, \bibinfo {author} {\bibfnamefont {S.~B.}\ \bibnamefont {Cahn}},
  \bibinfo {author} {\bibfnamefont {J.~A.}\ \bibnamefont {Nikkel}}, \bibinfo
  {author} {\bibfnamefont {W.~F.}\ \bibnamefont {Vinen}},\ and\ \bibinfo
  {author} {\bibfnamefont {D.~N.}\ \bibnamefont {McKinsey}},\ }\bibfield
  {title} {\bibinfo {title} {Visualization study of counterflow in superfluid
  using metastable helium molecules},\ }\href
  {https://doi.org/10.1103/physrevlett.105.045301} {\bibfield  {journal}
  {\bibinfo  {journal} {Phys. Rev. Lett.}\ }\textbf {\bibinfo {volume} {105}},\
  \bibinfo {pages} {045301} (\bibinfo {year} {2010})}\BibitemShut {NoStop}%
\bibitem [{\citenamefont {Barenghi}\ \emph
  {et~al.}(2023{\natexlab{b}})\citenamefont {Barenghi}, \citenamefont
  {Middleton-Spencer}, \citenamefont {Galantucci},\ and\ \citenamefont
  {Parker}}]{barenghi2023types}%
  \BibitemOpen
  \bibfield  {author} {\bibinfo {author} {\bibfnamefont {C.}~\bibnamefont
  {Barenghi}}, \bibinfo {author} {\bibfnamefont {H.}~\bibnamefont
  {Middleton-Spencer}}, \bibinfo {author} {\bibfnamefont {L.}~\bibnamefont
  {Galantucci}},\ and\ \bibinfo {author} {\bibfnamefont {N.}~\bibnamefont
  {Parker}},\ }\bibfield  {title} {\bibinfo {title} {Types of quantum
  turbulence},\ }\href {https://doi.org/10.1116/5.0146107} {\bibfield
  {journal} {\bibinfo  {journal} {AVS Quantum Sci.}\ }\textbf {\bibinfo
  {volume} {5}},\ \bibinfo {pages} {025601} (\bibinfo {year}
  {2023}{\natexlab{b}})}\BibitemShut {NoStop}%
\bibitem [{\citenamefont {Henn}\ \emph {et~al.}(2009)\citenamefont {Henn},
  \citenamefont {Seman}, \citenamefont {Roati}, \citenamefont {Magalhães},\
  and\ \citenamefont {Bagnato}}]{Henn2009}%
  \BibitemOpen
  \bibfield  {author} {\bibinfo {author} {\bibfnamefont {E.~A.~L.}\
  \bibnamefont {Henn}}, \bibinfo {author} {\bibfnamefont {J.~A.}\ \bibnamefont
  {Seman}}, \bibinfo {author} {\bibfnamefont {G.}~\bibnamefont {Roati}},
  \bibinfo {author} {\bibfnamefont {K.~M.~F.}\ \bibnamefont {Magalhães}},\
  and\ \bibinfo {author} {\bibfnamefont {V.~S.}\ \bibnamefont {Bagnato}},\
  }\bibfield  {title} {\bibinfo {title} {Emergence of turbulence in an
  oscillating {Bose-Einstein} condensate},\ }\href
  {https://doi.org/10.1103/physrevlett.103.045301} {\bibfield  {journal}
  {\bibinfo  {journal} {Phys. Rev. Lett.}\ }\textbf {\bibinfo {volume} {103}},\
  \bibinfo {pages} {045301} (\bibinfo {year} {2009})}\BibitemShut {NoStop}%
\bibitem [{\citenamefont {Barenghi}\ \emph {et~al.}(2014)\citenamefont
  {Barenghi}, \citenamefont {L’vov},\ and\ \citenamefont
  {Roche}}]{Barenghi2014}%
  \BibitemOpen
  \bibfield  {author} {\bibinfo {author} {\bibfnamefont {C.~F.}\ \bibnamefont
  {Barenghi}}, \bibinfo {author} {\bibfnamefont {V.~S.}\ \bibnamefont
  {L’vov}},\ and\ \bibinfo {author} {\bibfnamefont {P.-E.}\ \bibnamefont
  {Roche}},\ }\bibfield  {title} {\bibinfo {title} {Experimental, numerical,
  and analytical velocity spectra in turbulent quantum fluid},\ }\href
  {https://doi.org/10.1073/pnas.1312548111} {\bibfield  {journal} {\bibinfo
  {journal} {Proc. Natl. Acad. Sci.}\ }\textbf {\bibinfo {volume} {111}},\
  \bibinfo {pages} {4683} (\bibinfo {year} {2014})}\BibitemShut {NoStop}%
\bibitem [{\citenamefont {Frisch}(1995)}]{frisch1995turbulence}%
  \BibitemOpen
  \bibfield  {author} {\bibinfo {author} {\bibfnamefont {U.}~\bibnamefont
  {Frisch}},\ }\href@noop {} {\emph {\bibinfo {title} {Turbulence: the legacy
  of AN Kolmogorov}}}\ (\bibinfo  {publisher} {Cambridge university press},\
  \bibinfo {year} {1995})\BibitemShut {NoStop}%
\bibitem [{\citenamefont {Maurer}\ and\ \citenamefont
  {Tabeling}(1998)}]{Maurer1998}%
  \BibitemOpen
  \bibfield  {author} {\bibinfo {author} {\bibfnamefont {J.}~\bibnamefont
  {Maurer}}\ and\ \bibinfo {author} {\bibfnamefont {P.}~\bibnamefont
  {Tabeling}},\ }\bibfield  {title} {\bibinfo {title} {Local investigation of
  superfluid turbulence},\ }\href {https://doi.org/10.1209/epl/i1998-00314-9}
  {\bibfield  {journal} {\bibinfo  {journal} {Europhys. Lett.}\ }\textbf
  {\bibinfo {volume} {43}},\ \bibinfo {pages} {29} (\bibinfo {year}
  {1998})}\BibitemShut {NoStop}%
\bibitem [{\citenamefont {Nore}\ \emph {et~al.}(1997)\citenamefont {Nore},
  \citenamefont {Abid},\ and\ \citenamefont {Brachet}}]{Nore1997}%
  \BibitemOpen
  \bibfield  {author} {\bibinfo {author} {\bibfnamefont {C.}~\bibnamefont
  {Nore}}, \bibinfo {author} {\bibfnamefont {M.}~\bibnamefont {Abid}},\ and\
  \bibinfo {author} {\bibfnamefont {M.~E.}\ \bibnamefont {Brachet}},\
  }\bibfield  {title} {\bibinfo {title} {Kolmogorov turbulence in
  low-temperature superflows},\ }\href
  {https://doi.org/10.1103/PhysRevLett.78.3896} {\bibfield  {journal} {\bibinfo
   {journal} {Phys. Rev. Lett.}\ }\textbf {\bibinfo {volume} {78}},\ \bibinfo
  {pages} {3896} (\bibinfo {year} {1997})}\BibitemShut {NoStop}%
\bibitem [{\citenamefont {Araki}\ \emph {et~al.}(2002)\citenamefont {Araki},
  \citenamefont {Tsubota},\ and\ \citenamefont {Nemirovskii}}]{Araki2002}%
  \BibitemOpen
  \bibfield  {author} {\bibinfo {author} {\bibfnamefont {T.}~\bibnamefont
  {Araki}}, \bibinfo {author} {\bibfnamefont {M.}~\bibnamefont {Tsubota}},\
  and\ \bibinfo {author} {\bibfnamefont {S.~K.}\ \bibnamefont {Nemirovskii}},\
  }\bibfield  {title} {\bibinfo {title} {Energy spectrum of superfluid
  turbulence with no normal-fluid component},\ }\href
  {https://doi.org/10.1103/physrevlett.89.145301} {\bibfield  {journal}
  {\bibinfo  {journal} {Phys. Rev. Lett.}\ }\textbf {\bibinfo {volume} {89}},\
  \bibinfo {pages} {145301} (\bibinfo {year} {2002})}\BibitemShut {NoStop}%
\bibitem [{\citenamefont {Kobayashi}\ and\ \citenamefont
  {Tsubota}(2007)}]{Kobayashi2007}%
  \BibitemOpen
  \bibfield  {author} {\bibinfo {author} {\bibfnamefont {M.}~\bibnamefont
  {Kobayashi}}\ and\ \bibinfo {author} {\bibfnamefont {M.}~\bibnamefont
  {Tsubota}},\ }\bibfield  {title} {\bibinfo {title} {Quantum turbulence in a
  trapped {Bose-Einstein} condensate},\ }\href
  {https://doi.org/10.1103/physreva.76.045603} {\bibfield  {journal} {\bibinfo
  {journal} {Phys. Rev. A}\ }\textbf {\bibinfo {volume} {76}},\ \bibinfo
  {pages} {045603} (\bibinfo {year} {2007})}\BibitemShut {NoStop}%
\bibitem [{\citenamefont {Kobayashi}\ and\ \citenamefont
  {Tsubota}(2008)}]{Kobayashi2008}%
  \BibitemOpen
  \bibfield  {author} {\bibinfo {author} {\bibfnamefont {M.}~\bibnamefont
  {Kobayashi}}\ and\ \bibinfo {author} {\bibfnamefont {M.}~\bibnamefont
  {Tsubota}},\ }\bibfield  {title} {\bibinfo {title} {Quantum turbulence in a
  trapped {Bose-Einstein} condensate under combined rotations around three
  axes},\ }\href {https://doi.org/10.1007/s10909-007-9594-4} {\bibfield
  {journal} {\bibinfo  {journal} {J. Low Temp. Phys.}\ }\textbf {\bibinfo
  {volume} {150}},\ \bibinfo {pages} {587} (\bibinfo {year}
  {2008})}\BibitemShut {NoStop}%
\bibitem [{\citenamefont {Neely}\ \emph {et~al.}(2013)\citenamefont {Neely},
  \citenamefont {Bradley}, \citenamefont {Samson}, \citenamefont {Rooney},
  \citenamefont {Wright}, \citenamefont {Law}, \citenamefont
  {Carretero-González}, \citenamefont {Kevrekidis}, \citenamefont {Davis},\
  and\ \citenamefont {Anderson}}]{Neely2013}%
  \BibitemOpen
  \bibfield  {author} {\bibinfo {author} {\bibfnamefont {T.~W.}\ \bibnamefont
  {Neely}}, \bibinfo {author} {\bibfnamefont {A.~S.}\ \bibnamefont {Bradley}},
  \bibinfo {author} {\bibfnamefont {E.~C.}\ \bibnamefont {Samson}}, \bibinfo
  {author} {\bibfnamefont {S.~J.}\ \bibnamefont {Rooney}}, \bibinfo {author}
  {\bibfnamefont {E.~M.}\ \bibnamefont {Wright}}, \bibinfo {author}
  {\bibfnamefont {K.~J.~H.}\ \bibnamefont {Law}}, \bibinfo {author}
  {\bibfnamefont {R.}~\bibnamefont {Carretero-González}}, \bibinfo {author}
  {\bibfnamefont {P.~G.}\ \bibnamefont {Kevrekidis}}, \bibinfo {author}
  {\bibfnamefont {M.~J.}\ \bibnamefont {Davis}},\ and\ \bibinfo {author}
  {\bibfnamefont {B.~P.}\ \bibnamefont {Anderson}},\ }\bibfield  {title}
  {\bibinfo {title} {Characteristics of two-dimensional quantum turbulence in a
  compressible superfluid},\ }\href
  {https://doi.org/10.1103/physrevlett.111.235301} {\bibfield  {journal}
  {\bibinfo  {journal} {Phys. Rev. Lett.}\ }\textbf {\bibinfo {volume} {111}},\
  \bibinfo {pages} {235301} (\bibinfo {year} {2013})}\BibitemShut {NoStop}%
\bibitem [{\citenamefont {Wilson}\ \emph {et~al.}(2013)\citenamefont {Wilson},
  \citenamefont {Samson}, \citenamefont {Newman}, \citenamefont {Neely},\ and\
  \citenamefont {Anderson}}]{Wilson2013}%
  \BibitemOpen
  \bibfield  {author} {\bibinfo {author} {\bibfnamefont {K.~E.}\ \bibnamefont
  {Wilson}}, \bibinfo {author} {\bibfnamefont {E.~C.}\ \bibnamefont {Samson}},
  \bibinfo {author} {\bibfnamefont {Z.~L.}\ \bibnamefont {Newman}}, \bibinfo
  {author} {\bibfnamefont {T.~W.}\ \bibnamefont {Neely}},\ and\ \bibinfo
  {author} {\bibfnamefont {B.~P.}\ \bibnamefont {Anderson}},\ }\bibinfo {title}
  {Experimental methods for generating two-dimensional quantum turbulence in
  {Bose-Einstein} condensates},\ in\ \href
  {https://doi.org/10.1142/9789814440400_0007} {\emph {\bibinfo {booktitle}
  {Annual Review of Cold Atoms and Molecules}}}\ (\bibinfo  {publisher} {World
  Scientific},\ \bibinfo {year} {2013})\ pp.\ \bibinfo {pages}
  {261--298}\BibitemShut {NoStop}%
\bibitem [{\citenamefont {Numasato}\ \emph {et~al.}(2010)\citenamefont
  {Numasato}, \citenamefont {Tsubota},\ and\ \citenamefont
  {L’vov}}]{Numasato2010}%
  \BibitemOpen
  \bibfield  {author} {\bibinfo {author} {\bibfnamefont {R.}~\bibnamefont
  {Numasato}}, \bibinfo {author} {\bibfnamefont {M.}~\bibnamefont {Tsubota}},\
  and\ \bibinfo {author} {\bibfnamefont {V.~S.}\ \bibnamefont {L’vov}},\
  }\bibfield  {title} {\bibinfo {title} {Direct energy cascade in
  two-dimensional compressible quantum turbulence},\ }\href
  {https://doi.org/10.1103/physreva.81.063630} {\bibfield  {journal} {\bibinfo
  {journal} {Phys. Rev. A}\ }\textbf {\bibinfo {volume} {81}},\ \bibinfo
  {pages} {063630} (\bibinfo {year} {2010})}\BibitemShut {NoStop}%
\bibitem [{\citenamefont {Müller}\ \emph {et~al.}(2020)\citenamefont
  {Müller}, \citenamefont {Brachet}, \citenamefont {Alexakis},\ and\
  \citenamefont {Mininni}}]{Mueller2020}%
  \BibitemOpen
  \bibfield  {author} {\bibinfo {author} {\bibfnamefont {N.~P.}\ \bibnamefont
  {Müller}}, \bibinfo {author} {\bibfnamefont {M.-E.}\ \bibnamefont
  {Brachet}}, \bibinfo {author} {\bibfnamefont {A.}~\bibnamefont {Alexakis}},\
  and\ \bibinfo {author} {\bibfnamefont {P.~D.}\ \bibnamefont {Mininni}},\
  }\bibfield  {title} {\bibinfo {title} {Abrupt transition between
  three-dimensional and two-dimensional quantum turbulence},\ }\href
  {https://doi.org/10.1103/physrevlett.124.134501} {\bibfield  {journal}
  {\bibinfo  {journal} {Phys. Rev. Lett.}\ }\textbf {\bibinfo {volume} {124}},\
  \bibinfo {pages} {134501} (\bibinfo {year} {2020})}\BibitemShut {NoStop}%
\bibitem [{\citenamefont {Reeves}\ \emph {et~al.}(2013)\citenamefont {Reeves},
  \citenamefont {Billam}, \citenamefont {Anderson},\ and\ \citenamefont
  {Bradley}}]{Reeves2013}%
  \BibitemOpen
  \bibfield  {author} {\bibinfo {author} {\bibfnamefont {M.~T.}\ \bibnamefont
  {Reeves}}, \bibinfo {author} {\bibfnamefont {T.~P.}\ \bibnamefont {Billam}},
  \bibinfo {author} {\bibfnamefont {B.~P.}\ \bibnamefont {Anderson}},\ and\
  \bibinfo {author} {\bibfnamefont {A.~S.}\ \bibnamefont {Bradley}},\
  }\bibfield  {title} {\bibinfo {title} {Inverse energy cascade in forced
  two-dimensional quantum turbulence},\ }\href
  {https://doi.org/10.1103/physrevlett.110.104501} {\bibfield  {journal}
  {\bibinfo  {journal} {Phys. Rev. Lett.}\ }\textbf {\bibinfo {volume} {110}},\
  \bibinfo {pages} {104501} (\bibinfo {year} {2013})}\BibitemShut {NoStop}%
\bibitem [{\citenamefont {Volovik}(2004)}]{Volovik2004}%
  \BibitemOpen
  \bibfield  {author} {\bibinfo {author} {\bibfnamefont {G.~E.}\ \bibnamefont
  {Volovik}},\ }\bibfield  {title} {\bibinfo {title} {On developed superfluid
  turbulence},\ }\href {https://doi.org/10.1023/b:jolt.0000041269.56070.2d}
  {\bibfield  {journal} {\bibinfo  {journal} {J. Low Temp. Phys.}\ }\textbf
  {\bibinfo {volume} {136}},\ \bibinfo {pages} {309} (\bibinfo {year}
  {2004})}\BibitemShut {NoStop}%
\bibitem [{\citenamefont {Bradley}\ \emph {et~al.}(2006)\citenamefont
  {Bradley}, \citenamefont {Clubb}, \citenamefont {Fisher}, \citenamefont
  {Guénault}, \citenamefont {Haley}, \citenamefont {Matthews}, \citenamefont
  {Pickett}, \citenamefont {Tsepelin},\ and\ \citenamefont
  {Zaki}}]{Bradley2006}%
  \BibitemOpen
  \bibfield  {author} {\bibinfo {author} {\bibfnamefont {D.~I.}\ \bibnamefont
  {Bradley}}, \bibinfo {author} {\bibfnamefont {D.~O.}\ \bibnamefont {Clubb}},
  \bibinfo {author} {\bibfnamefont {S.~N.}\ \bibnamefont {Fisher}}, \bibinfo
  {author} {\bibfnamefont {A.~M.}\ \bibnamefont {Guénault}}, \bibinfo {author}
  {\bibfnamefont {R.~P.}\ \bibnamefont {Haley}}, \bibinfo {author}
  {\bibfnamefont {C.~J.}\ \bibnamefont {Matthews}}, \bibinfo {author}
  {\bibfnamefont {G.~R.}\ \bibnamefont {Pickett}}, \bibinfo {author}
  {\bibfnamefont {V.}~\bibnamefont {Tsepelin}},\ and\ \bibinfo {author}
  {\bibfnamefont {K.}~\bibnamefont {Zaki}},\ }\bibfield  {title} {\bibinfo
  {title} {Decay of pure quantum turbulence in superfluid},\ }\href
  {https://doi.org/10.1103/physrevlett.96.035301} {\bibfield  {journal}
  {\bibinfo  {journal} {Phys. Rev. Lett.}\ }\textbf {\bibinfo {volume} {96}},\
  \bibinfo {pages} {035301} (\bibinfo {year} {2006})}\BibitemShut {NoStop}%
\bibitem [{\citenamefont {Cidrim}\ \emph {et~al.}(2017)\citenamefont {Cidrim},
  \citenamefont {White}, \citenamefont {Allen}, \citenamefont {Bagnato},\ and\
  \citenamefont {Barenghi}}]{Cidrim2017}%
  \BibitemOpen
  \bibfield  {author} {\bibinfo {author} {\bibfnamefont {A.}~\bibnamefont
  {Cidrim}}, \bibinfo {author} {\bibfnamefont {A.~C.}\ \bibnamefont {White}},
  \bibinfo {author} {\bibfnamefont {A.~J.}\ \bibnamefont {Allen}}, \bibinfo
  {author} {\bibfnamefont {V.~S.}\ \bibnamefont {Bagnato}},\ and\ \bibinfo
  {author} {\bibfnamefont {C.~F.}\ \bibnamefont {Barenghi}},\ }\bibfield
  {title} {\bibinfo {title} {Vinen turbulence via the decay of multicharged
  vortices in trapped atomic {Bose-Einstein condensates}},\ }\href
  {https://doi.org/10.1103/physreva.96.023617} {\bibfield  {journal} {\bibinfo
  {journal} {Phys. Rev. A}\ }\textbf {\bibinfo {volume} {96}},\ \bibinfo
  {pages} {023617} (\bibinfo {year} {2017})}\BibitemShut {NoStop}%
\bibitem [{\citenamefont {Tsubota}\ \emph {et~al.}(2013)\citenamefont
  {Tsubota}, \citenamefont {Kobayashi},\ and\ \citenamefont
  {Takeuchi}}]{Tsubota2013}%
  \BibitemOpen
  \bibfield  {author} {\bibinfo {author} {\bibfnamefont {M.}~\bibnamefont
  {Tsubota}}, \bibinfo {author} {\bibfnamefont {M.}~\bibnamefont {Kobayashi}},\
  and\ \bibinfo {author} {\bibfnamefont {H.}~\bibnamefont {Takeuchi}},\
  }\bibfield  {title} {\bibinfo {title} {Quantum hydrodynamics},\ }\href
  {https://doi.org/10.1016/j.physrep.2012.09.007} {\bibfield  {journal}
  {\bibinfo  {journal} {Phys. Rep.}\ }\textbf {\bibinfo {volume} {522}},\
  \bibinfo {pages} {191} (\bibinfo {year} {2013})}\BibitemShut {NoStop}%
\bibitem [{\citenamefont {Middleton-Spencer}\ \emph {et~al.}(2023)\citenamefont
  {Middleton-Spencer}, \citenamefont {Orozco}, \citenamefont {Galantucci},
  \citenamefont {Moreno}, \citenamefont {Parker}, \citenamefont {Machado},
  \citenamefont {Bagnato},\ and\ \citenamefont
  {Barenghi}}]{MiddletonSpencer2023}%
  \BibitemOpen
  \bibfield  {author} {\bibinfo {author} {\bibfnamefont {H.~A.~J.}\
  \bibnamefont {Middleton-Spencer}}, \bibinfo {author} {\bibfnamefont
  {A.~D.~G.}\ \bibnamefont {Orozco}}, \bibinfo {author} {\bibfnamefont
  {L.}~\bibnamefont {Galantucci}}, \bibinfo {author} {\bibfnamefont
  {M.}~\bibnamefont {Moreno}}, \bibinfo {author} {\bibfnamefont {N.~G.}\
  \bibnamefont {Parker}}, \bibinfo {author} {\bibfnamefont {L.~A.}\
  \bibnamefont {Machado}}, \bibinfo {author} {\bibfnamefont {V.~S.}\
  \bibnamefont {Bagnato}},\ and\ \bibinfo {author} {\bibfnamefont {C.~F.}\
  \bibnamefont {Barenghi}},\ }\bibfield  {title} {\bibinfo {title} {Strong
  quantum turbulence in {Bose-Einstein} condensates},\ }\href
  {https://doi.org/10.1103/physrevresearch.5.043081} {\bibfield  {journal}
  {\bibinfo  {journal} {Phys. Rev. Res.}\ }\textbf {\bibinfo {volume} {5}},\
  \bibinfo {pages} {043081} (\bibinfo {year} {2023})}\BibitemShut {NoStop}%
\bibitem [{\citenamefont {Navon}\ \emph {et~al.}(2016)\citenamefont {Navon},
  \citenamefont {Gaunt}, \citenamefont {Smith},\ and\ \citenamefont
  {Hadzibabic}}]{Navon2016}%
  \BibitemOpen
  \bibfield  {author} {\bibinfo {author} {\bibfnamefont {N.}~\bibnamefont
  {Navon}}, \bibinfo {author} {\bibfnamefont {A.~L.}\ \bibnamefont {Gaunt}},
  \bibinfo {author} {\bibfnamefont {R.~P.}\ \bibnamefont {Smith}},\ and\
  \bibinfo {author} {\bibfnamefont {Z.}~\bibnamefont {Hadzibabic}},\ }\bibfield
   {title} {\bibinfo {title} {Emergence of a turbulent cascade in a quantum
  gas},\ }\href {https://doi.org/10.1038/nature20114} {\bibfield  {journal}
  {\bibinfo  {journal} {Nature (London)}\ }\textbf {\bibinfo {volume} {539}},\
  \bibinfo {pages} {72} (\bibinfo {year} {2016})}\BibitemShut {NoStop}%
\bibitem [{\citenamefont {Sivakumar}\ \emph
  {et~al.}(2024{\natexlab{a}})\citenamefont {Sivakumar}, \citenamefont
  {Mishra}, \citenamefont {Hujeirat},\ and\ \citenamefont
  {Muruganandam}}]{Sivakumar2024a}%
  \BibitemOpen
  \bibfield  {author} {\bibinfo {author} {\bibfnamefont {A.}~\bibnamefont
  {Sivakumar}}, \bibinfo {author} {\bibfnamefont {P.~K.}\ \bibnamefont
  {Mishra}}, \bibinfo {author} {\bibfnamefont {A.~A.}\ \bibnamefont
  {Hujeirat}},\ and\ \bibinfo {author} {\bibfnamefont {P.}~\bibnamefont
  {Muruganandam}},\ }\bibfield  {title} {\bibinfo {title} {{Dynamic
  instabilities and turbulence of merged rotating Bose-Einstein condensates}},\
  }\href {https://doi.org/10.1063/5.0231764} {\bibfield  {journal} {\bibinfo
  {journal} {Phys. Fluids}\ }\textbf {\bibinfo {volume} {36}},\ \bibinfo
  {pages} {117121} (\bibinfo {year} {2024}{\natexlab{a}})}\BibitemShut
  {NoStop}%
\bibitem [{\citenamefont {Estrada}\ \emph
  {et~al.}(2022{\natexlab{a}})\citenamefont {Estrada}, \citenamefont
  {Brachet},\ and\ \citenamefont {Mininni}}]{AmetteEstrada2022}%
  \BibitemOpen
  \bibfield  {author} {\bibinfo {author} {\bibfnamefont {J.~A.}\ \bibnamefont
  {Estrada}}, \bibinfo {author} {\bibfnamefont {M.~E.}\ \bibnamefont
  {Brachet}},\ and\ \bibinfo {author} {\bibfnamefont {P.~D.}\ \bibnamefont
  {Mininni}},\ }\bibfield  {title} {\bibinfo {title} {Turbulence in rotating
  {Bose-Einstein} condensates},\ }\href
  {https://doi.org/10.1103/PhysRevA.105.063321} {\bibfield  {journal} {\bibinfo
   {journal} {Phys. Rev. A}\ }\textbf {\bibinfo {volume} {105}},\ \bibinfo
  {pages} {063321} (\bibinfo {year} {2022}{\natexlab{a}})}\BibitemShut
  {NoStop}%
\bibitem [{\citenamefont {Estrada}\ \emph
  {et~al.}(2022{\natexlab{b}})\citenamefont {Estrada}, \citenamefont
  {Brachet},\ and\ \citenamefont {Mininni}}]{Estrada2022}%
  \BibitemOpen
  \bibfield  {author} {\bibinfo {author} {\bibfnamefont {J.~A.}\ \bibnamefont
  {Estrada}}, \bibinfo {author} {\bibfnamefont {M.~E.}\ \bibnamefont
  {Brachet}},\ and\ \bibinfo {author} {\bibfnamefont {P.~D.}\ \bibnamefont
  {Mininni}},\ }\bibfield  {title} {\bibinfo {title} {Thermalized abrikosov
  lattices from decaying turbulence in rotating becs},\ }\href
  {https://doi.org/10.1116/5.0123277} {\bibfield  {journal} {\bibinfo
  {journal} {AVS Quantum Sci.}\ }\textbf {\bibinfo {volume} {4}},\ \bibinfo
  {pages} {046201} (\bibinfo {year} {2022}{\natexlab{b}})}\BibitemShut
  {NoStop}%
\bibitem [{\citenamefont {Sivakumar}\ \emph
  {et~al.}(2024{\natexlab{b}})\citenamefont {Sivakumar}, \citenamefont
  {Mishra}, \citenamefont {Hujeirat},\ and\ \citenamefont
  {Muruganandam}}]{Sivakumar2024}%
  \BibitemOpen
  \bibfield  {author} {\bibinfo {author} {\bibfnamefont {A.}~\bibnamefont
  {Sivakumar}}, \bibinfo {author} {\bibfnamefont {P.~K.}\ \bibnamefont
  {Mishra}}, \bibinfo {author} {\bibfnamefont {A.~A.}\ \bibnamefont
  {Hujeirat}},\ and\ \bibinfo {author} {\bibfnamefont {P.}~\bibnamefont
  {Muruganandam}},\ }\bibfield  {title} {\bibinfo {title} {Energy spectra and
  fluxes of turbulent rotating {Bose–Einstein} condensates in two
  dimensions},\ }\href {https://doi.org/10.1063/5.0190917} {\bibfield
  {journal} {\bibinfo  {journal} {Phys. Fluids}\ }\textbf {\bibinfo {volume}
  {36}},\ \bibinfo {pages} {027149} (\bibinfo {year}
  {2024}{\natexlab{b}})}\BibitemShut {NoStop}%
\bibitem [{\citenamefont {Haskell}\ and\ \citenamefont
  {Melatos}(2015)}]{Haskell2015}%
  \BibitemOpen
  \bibfield  {author} {\bibinfo {author} {\bibfnamefont {B.}~\bibnamefont
  {Haskell}}\ and\ \bibinfo {author} {\bibfnamefont {A.}~\bibnamefont
  {Melatos}},\ }\bibfield  {title} {\bibinfo {title} {Models of pulsar
  glitches},\ }\href {https://doi.org/10.1142/s0218271815300086} {\bibfield
  {journal} {\bibinfo  {journal} {Int. J. Mod. Phys. D}\ }\textbf {\bibinfo
  {volume} {24}},\ \bibinfo {pages} {1530008} (\bibinfo {year}
  {2015})}\BibitemShut {NoStop}%
\bibitem [{\citenamefont {Hui}\ \emph {et~al.}(2017)\citenamefont {Hui},
  \citenamefont {Ostriker}, \citenamefont {Tremaine},\ and\ \citenamefont
  {Witten}}]{Hui2017}%
  \BibitemOpen
  \bibfield  {author} {\bibinfo {author} {\bibfnamefont {L.}~\bibnamefont
  {Hui}}, \bibinfo {author} {\bibfnamefont {J.~P.}\ \bibnamefont {Ostriker}},
  \bibinfo {author} {\bibfnamefont {S.}~\bibnamefont {Tremaine}},\ and\
  \bibinfo {author} {\bibfnamefont {E.}~\bibnamefont {Witten}},\ }\bibfield
  {title} {\bibinfo {title} {Ultralight scalars as cosmological dark matter},\
  }\href {https://doi.org/10.1103/physrevd.95.043541} {\bibfield  {journal}
  {\bibinfo  {journal} {Phys. Rev. D}\ }\textbf {\bibinfo {volume} {95}},\
  \bibinfo {pages} {043541} (\bibinfo {year} {2017})}\BibitemShut {NoStop}%
\bibitem [{\citenamefont {Chavanis}\ and\ \citenamefont
  {Harko}(2012)}]{Chavanis2012}%
  \BibitemOpen
  \bibfield  {author} {\bibinfo {author} {\bibfnamefont {P.-H.}\ \bibnamefont
  {Chavanis}}\ and\ \bibinfo {author} {\bibfnamefont {T.}~\bibnamefont
  {Harko}},\ }\bibfield  {title} {\bibinfo {title} {{Bose-Einstein} condensate
  general relativistic stars},\ }\href
  {https://doi.org/10.1103/physrevd.86.064011} {\bibfield  {journal} {\bibinfo
  {journal} {Phys. Rev. D}\ }\textbf {\bibinfo {volume} {86}},\ \bibinfo
  {pages} {064011} (\bibinfo {year} {2012})}\BibitemShut {NoStop}%
\bibitem [{\citenamefont {Chavanis}(2014)}]{Chavanis2014}%
  \BibitemOpen
  \bibfield  {author} {\bibinfo {author} {\bibfnamefont {P.-H.}\ \bibnamefont
  {Chavanis}},\ }\bibinfo {title} {Self-gravitating {Bose-Einstein}
  condensates},\ in\ \href {https://doi.org/10.1007/978-3-319-10852-0_6} {\emph
  {\bibinfo {booktitle} {Quantum Aspects of Black Holes}}}\ (\bibinfo
  {publisher} {Springer International Publishing},\ \bibinfo {year} {2014})\
  pp.\ \bibinfo {pages} {151--194}\BibitemShut {NoStop}%
\bibitem [{\citenamefont {Ferreira}(2021)}]{Ferreira2021}%
  \BibitemOpen
  \bibfield  {author} {\bibinfo {author} {\bibfnamefont {E.~G.~M.}\
  \bibnamefont {Ferreira}},\ }\bibfield  {title} {\bibinfo {title} {Ultra-light
  dark matter},\ }\href {https://doi.org/10.1007/s00159-021-00135-6} {\bibfield
   {journal} {\bibinfo  {journal} {Astron. Astrophys. Rev.}\ }\textbf {\bibinfo
  {volume} {29}},\ \bibinfo {pages} {7} (\bibinfo {year} {2021})}\BibitemShut
  {NoStop}%
\bibitem [{\citenamefont {Böhmer}\ and\ \citenamefont
  {Harko}(2007)}]{Boehmer2007}%
  \BibitemOpen
  \bibfield  {author} {\bibinfo {author} {\bibfnamefont {C.~G.}\ \bibnamefont
  {Böhmer}}\ and\ \bibinfo {author} {\bibfnamefont {T.}~\bibnamefont
  {Harko}},\ }\bibfield  {title} {\bibinfo {title} {Can dark matter be a
  bose–einstein condensate?},\ }\href
  {https://doi.org/10.1088/1475-7516/2007/06/025} {\bibfield  {journal}
  {\bibinfo  {journal} {J. Cosmol. Astropart. P.}\ }\textbf {\bibinfo {volume}
  {2007}},\ \bibinfo {pages} {025} (\bibinfo {year} {2007})}\BibitemShut
  {NoStop}%
\bibitem [{\citenamefont {Shukla}(2025)}]{shukla_2025_GPPE_turbulence}%
  \BibitemOpen
  \bibfield  {author} {\bibinfo {author} {\bibfnamefont {S.}~\bibnamefont
  {Shukla}},\ }\bibfield  {title} {\bibinfo {title} {Turbulence and large-scale
  structures in self-gravitating superfluids},\ }\href
  {https://doi.org/10.1103/7jtd-kybt} {\bibfield  {journal} {\bibinfo
  {journal} {Phys. Rev. Fluids}\ }\textbf {\bibinfo {volume} {10}},\ \bibinfo
  {pages} {114601} (\bibinfo {year} {2025})}\BibitemShut {NoStop}%
\bibitem [{\citenamefont {Mocz}\ \emph {et~al.}(2017)\citenamefont {Mocz},
  \citenamefont {Vogelsberger}, \citenamefont {Robles}, \citenamefont {Zavala},
  \citenamefont {Boylan-Kolchin}, \citenamefont {Fialkov},\ and\ \citenamefont
  {Hernquist}}]{Mocz2017}%
  \BibitemOpen
  \bibfield  {author} {\bibinfo {author} {\bibfnamefont {P.}~\bibnamefont
  {Mocz}}, \bibinfo {author} {\bibfnamefont {M.}~\bibnamefont {Vogelsberger}},
  \bibinfo {author} {\bibfnamefont {V.~H.}\ \bibnamefont {Robles}}, \bibinfo
  {author} {\bibfnamefont {J.}~\bibnamefont {Zavala}}, \bibinfo {author}
  {\bibfnamefont {M.}~\bibnamefont {Boylan-Kolchin}}, \bibinfo {author}
  {\bibfnamefont {A.}~\bibnamefont {Fialkov}},\ and\ \bibinfo {author}
  {\bibfnamefont {L.}~\bibnamefont {Hernquist}},\ }\bibfield  {title} {\bibinfo
  {title} {Galaxy formation with {BECDM - I}. {T}urbulence and relaxation of
  idealized haloes},\ }\href {https://doi.org/10.1093/mnras/stx1887} {\bibfield
   {journal} {\bibinfo  {journal} {Mon. Not. R. Astron. Soc.}\ }\textbf
  {\bibinfo {volume} {471}},\ \bibinfo {pages} {4559} (\bibinfo {year}
  {2017})}\BibitemShut {NoStop}%
\bibitem [{\citenamefont {Sivakumar}\ \emph {et~al.}(2025)\citenamefont
  {Sivakumar}, \citenamefont {Mishra}, \citenamefont {Hujeirat},\ and\
  \citenamefont {Muruganandam}}]{Sivakumar2025}%
  \BibitemOpen
  \bibfield  {author} {\bibinfo {author} {\bibfnamefont {A.}~\bibnamefont
  {Sivakumar}}, \bibinfo {author} {\bibfnamefont {P.~K.}\ \bibnamefont
  {Mishra}}, \bibinfo {author} {\bibfnamefont {A.~A.}\ \bibnamefont
  {Hujeirat}},\ and\ \bibinfo {author} {\bibfnamefont {P.}~\bibnamefont
  {Muruganandam}},\ }\bibfield  {title} {\bibinfo {title} {Revealing turbulent
  dark matter via merging of self-gravitating condensates},\ }\href
  {https://doi.org/10.1103/physrevd.111.083511} {\bibfield  {journal} {\bibinfo
   {journal} {Phys. Rev. D}\ }\textbf {\bibinfo {volume} {111}},\ \bibinfo
  {pages} {083511} (\bibinfo {year} {2025})}\BibitemShut {NoStop}%
\bibitem [{\citenamefont {Radhakrishnan}\ and\ \citenamefont
  {Manchester}(1969)}]{Radhakrishnan1969}%
  \BibitemOpen
  \bibfield  {author} {\bibinfo {author} {\bibfnamefont {V.}~\bibnamefont
  {Radhakrishnan}}\ and\ \bibinfo {author} {\bibfnamefont {R.~N.}\ \bibnamefont
  {Manchester}},\ }\bibfield  {title} {\bibinfo {title} {Detection of a change
  of state in the pulsar {PSR 0833-45}},\ }\href
  {https://doi.org/10.1038/222228a0} {\bibfield  {journal} {\bibinfo  {journal}
  {Nature (London)}\ }\textbf {\bibinfo {volume} {222}},\ \bibinfo {pages}
  {228} (\bibinfo {year} {1969})}\BibitemShut {NoStop}%
\bibitem [{\citenamefont {Boynton}\ \emph {et~al.}(1969)\citenamefont
  {Boynton}, \citenamefont {{Groth III}}, \citenamefont {Partridge},\ and\
  \citenamefont {Wilkinson}}]{Boynton1969}%
  \BibitemOpen
  \bibfield  {author} {\bibinfo {author} {\bibfnamefont {P.~E.}\ \bibnamefont
  {Boynton}}, \bibinfo {author} {\bibfnamefont {E.~J.}\ \bibnamefont {{Groth
  III}}}, \bibinfo {author} {\bibfnamefont {B.}~\bibnamefont {Partridge}},\
  and\ \bibinfo {author} {\bibfnamefont {D.~T.}\ \bibnamefont {Wilkinson}},\
  }\bibfield  {title} {\bibinfo {title} {Precision measurement of the frequency
  decay of the {Crab Nebula Pulsar, NP 0532}},\ }\href@noop {} {\bibfield
  {journal} {\bibinfo  {journal} {Astrophys. J.}\ }\textbf {\bibinfo {volume}
  {157}},\ \bibinfo {pages} {L197} (\bibinfo {year} {1969})}\BibitemShut
  {NoStop}%
\bibitem [{\citenamefont {Manchester}(2017)}]{Manchester2017}%
  \BibitemOpen
  \bibfield  {author} {\bibinfo {author} {\bibfnamefont {R.~N.}\ \bibnamefont
  {Manchester}},\ }\bibfield  {title} {\bibinfo {title} {Pulsar glitches},\
  }\href {https://doi.org/10.1017/S1743921317009607} {\bibfield  {journal}
  {\bibinfo  {journal} {IAU Symp.}\ }\textbf {\bibinfo {volume} {337}},\
  \bibinfo {pages} {197} (\bibinfo {year} {2017})}\BibitemShut {NoStop}%
\bibitem [{\citenamefont {Jones}(1986)}]{Jones1986}%
  \BibitemOpen
  \bibfield  {author} {\bibinfo {author} {\bibfnamefont {P.~B.}\ \bibnamefont
  {Jones}},\ }\bibfield  {title} {\bibinfo {title} {Properties of condensed
  matter in very strong magnetic fields},\ }\href
  {https://doi.org/10.1093/mnras/218.3.477} {\bibfield  {journal} {\bibinfo
  {journal} {Mon. Not. R. Astron. Soc.}\ }\textbf {\bibinfo {volume} {218}},\
  \bibinfo {pages} {477} (\bibinfo {year} {1986})}\BibitemShut {NoStop}%
\bibitem [{\citenamefont {Khomenko}\ and\ \citenamefont
  {Haskell}(2018)}]{Khomenko2018}%
  \BibitemOpen
  \bibfield  {author} {\bibinfo {author} {\bibfnamefont {V.}~\bibnamefont
  {Khomenko}}\ and\ \bibinfo {author} {\bibfnamefont {B.}~\bibnamefont
  {Haskell}},\ }\bibfield  {title} {\bibinfo {title} {Modelling pulsar
  glitches: {The} hydrodynamics of superfluid vortex avalanches in neutron
  stars},\ }\href {https://doi.org/10.1017/pasa.2018.12} {\bibfield  {journal}
  {\bibinfo  {journal} {Pub. Astron. Soc. Austrl.}\ }\textbf {\bibinfo {volume}
  {35}},\ \bibinfo {pages} {e020} (\bibinfo {year} {2018})}\BibitemShut
  {NoStop}%
\bibitem [{\citenamefont {Howitt}\ and\ \citenamefont
  {Melatos}(2022)}]{Howitt2022}%
  \BibitemOpen
  \bibfield  {author} {\bibinfo {author} {\bibfnamefont {G.}~\bibnamefont
  {Howitt}}\ and\ \bibinfo {author} {\bibfnamefont {A.}~\bibnamefont
  {Melatos}},\ }\bibfield  {title} {\bibinfo {title} {Antiglitches in accreting
  pulsars from superfluid vortex avalanches},\ }\href
  {https://doi.org/10.1093/mnras/stac1358} {\bibfield  {journal} {\bibinfo
  {journal} {Mon. Not. R. Astron. Soc.}\ }\textbf {\bibinfo {volume} {514}},\
  \bibinfo {pages} {863} (\bibinfo {year} {2022})}\BibitemShut {NoStop}%
\bibitem [{\citenamefont {Tsubota}\ and\ \citenamefont
  {Maekawa}(1993)}]{Tsubota1993}%
  \BibitemOpen
  \bibfield  {author} {\bibinfo {author} {\bibfnamefont {M.}~\bibnamefont
  {Tsubota}}\ and\ \bibinfo {author} {\bibfnamefont {S.}~\bibnamefont
  {Maekawa}},\ }\bibfield  {title} {\bibinfo {title} {Pinning and depinning of
  two quantized vortices in superfluid \(^4\){He}},\ }\href
  {https://doi.org/10.1103/physrevb.47.12040} {\bibfield  {journal} {\bibinfo
  {journal} {Phys. Rev. B}\ }\textbf {\bibinfo {volume} {47}},\ \bibinfo
  {pages} {12040} (\bibinfo {year} {1993})}\BibitemShut {NoStop}%
\bibitem [{\citenamefont {Amico}\ \emph {et~al.}(2021)\citenamefont {Amico},
  \citenamefont {Boshier}, \citenamefont {Birkl}, \citenamefont {Minguzzi},
  \citenamefont {Miniatura}, \citenamefont {Kwek}, \citenamefont {Aghamalyan},
  \citenamefont {Ahufinger}, \citenamefont {Anderson}, \citenamefont {Andrei}
  \emph {et~al.}}]{Amico2021}%
  \BibitemOpen
  \bibfield  {author} {\bibinfo {author} {\bibfnamefont {L.}~\bibnamefont
  {Amico}}, \bibinfo {author} {\bibfnamefont {M.}~\bibnamefont {Boshier}},
  \bibinfo {author} {\bibfnamefont {G.}~\bibnamefont {Birkl}}, \bibinfo
  {author} {\bibfnamefont {A.}~\bibnamefont {Minguzzi}}, \bibinfo {author}
  {\bibfnamefont {C.}~\bibnamefont {Miniatura}}, \bibinfo {author}
  {\bibfnamefont {L.-C.}\ \bibnamefont {Kwek}}, \bibinfo {author}
  {\bibfnamefont {D.}~\bibnamefont {Aghamalyan}}, \bibinfo {author}
  {\bibfnamefont {V.}~\bibnamefont {Ahufinger}}, \bibinfo {author}
  {\bibfnamefont {D.}~\bibnamefont {Anderson}}, \bibinfo {author}
  {\bibfnamefont {N.}~\bibnamefont {Andrei}}, \emph {et~al.},\ }\bibfield
  {title} {\bibinfo {title} {Roadmap on atomtronics: State of the art and
  perspective},\ }\href {https://doi.org/10.1116/5.0026178} {\bibfield
  {journal} {\bibinfo  {journal} {AVS Quantum Sci.}\ }\textbf {\bibinfo
  {volume} {3}},\ \bibinfo {pages} {039201} (\bibinfo {year}
  {2021})}\BibitemShut {NoStop}%
\bibitem [{\citenamefont {Nelson}\ and\ \citenamefont
  {Vinokur}(1993)}]{Nelson1993}%
  \BibitemOpen
  \bibfield  {author} {\bibinfo {author} {\bibfnamefont {D.~R.}\ \bibnamefont
  {Nelson}}\ and\ \bibinfo {author} {\bibfnamefont {V.~M.}\ \bibnamefont
  {Vinokur}},\ }\bibfield  {title} {\bibinfo {title} {Boson localization and
  correlated pinning of superconducting vortex arrays},\ }\href
  {https://doi.org/10.1103/physrevb.48.13060} {\bibfield  {journal} {\bibinfo
  {journal} {Phys. Rev. B}\ }\textbf {\bibinfo {volume} {48}},\ \bibinfo
  {pages} {13060} (\bibinfo {year} {1993})}\BibitemShut {NoStop}%
\bibitem [{\citenamefont {Melatos}\ \emph {et~al.}(2008)\citenamefont
  {Melatos}, \citenamefont {Peralta},\ and\ \citenamefont
  {Wyithe}}]{Melatos2008}%
  \BibitemOpen
  \bibfield  {author} {\bibinfo {author} {\bibfnamefont {A.}~\bibnamefont
  {Melatos}}, \bibinfo {author} {\bibfnamefont {C.}~\bibnamefont {Peralta}},\
  and\ \bibinfo {author} {\bibfnamefont {J.~S.~B.}\ \bibnamefont {Wyithe}},\
  }\bibfield  {title} {\bibinfo {title} {Avalanche dynamics of radio pulsar
  glitches},\ }\href {https://doi.org/10.1086/523349} {\bibfield  {journal}
  {\bibinfo  {journal} {Astrophys. J.}\ }\textbf {\bibinfo {volume} {672}},\
  \bibinfo {pages} {1103} (\bibinfo {year} {2008})}\BibitemShut {NoStop}%
\bibitem [{\citenamefont {Lönnborn}\ \emph {et~al.}(2019)\citenamefont
  {Lönnborn}, \citenamefont {Melatos},\ and\ \citenamefont
  {Haskell}}]{Loennborn2019}%
  \BibitemOpen
  \bibfield  {author} {\bibinfo {author} {\bibfnamefont {J.~R.}\ \bibnamefont
  {Lönnborn}}, \bibinfo {author} {\bibfnamefont {A.}~\bibnamefont {Melatos}},\
  and\ \bibinfo {author} {\bibfnamefont {B.}~\bibnamefont {Haskell}},\
  }\bibfield  {title} {\bibinfo {title} {Collective, glitch-like vortex motion
  in a neutron star with an annular pinning barrier},\ }\href
  {https://doi.org/10.1093/mnras/stz1302} {\bibfield  {journal} {\bibinfo
  {journal} {Mon. Not. R. Astron. Soc.}\ }\textbf {\bibinfo {volume} {487}},\
  \bibinfo {pages} {702} (\bibinfo {year} {2019})}\BibitemShut {NoStop}%
\bibitem [{\citenamefont {Sonin}(1997)}]{Sonin1997}%
  \BibitemOpen
  \bibfield  {author} {\bibinfo {author} {\bibfnamefont {E.~B.}\ \bibnamefont
  {Sonin}},\ }\bibfield  {title} {\bibinfo {title} {Magnus force in superfluids
  and superconductors},\ }\href {https://doi.org/10.1103/PhysRevD.55.485}
  {\bibfield  {journal} {\bibinfo  {journal} {Phys. Rev. B}\ }\textbf {\bibinfo
  {volume} {55}},\ \bibinfo {pages} {485} (\bibinfo {year} {1997})}\BibitemShut
  {NoStop}%
\bibitem [{\citenamefont {Stockdale}\ \emph {et~al.}(2021)\citenamefont
  {Stockdale}, \citenamefont {Reeves},\ and\ \citenamefont
  {Davis}}]{Stockdale2021}%
  \BibitemOpen
  \bibfield  {author} {\bibinfo {author} {\bibfnamefont {O.~R.}\ \bibnamefont
  {Stockdale}}, \bibinfo {author} {\bibfnamefont {M.~T.}\ \bibnamefont
  {Reeves}},\ and\ \bibinfo {author} {\bibfnamefont {M.~J.}\ \bibnamefont
  {Davis}},\ }\bibfield  {title} {\bibinfo {title} {Dynamical mechanisms of
  vortex pinning in superfluid thin films},\ }\href
  {https://doi.org/10.1103/PhysRevLett.127.255302} {\bibfield  {journal}
  {\bibinfo  {journal} {Phys. Rev. Lett.}\ }\textbf {\bibinfo {volume} {127}},\
  \bibinfo {pages} {255302} (\bibinfo {year} {2021})}\BibitemShut {NoStop}%
\bibitem [{\citenamefont {Schwarz}(1981)}]{Schwarz1981}%
  \BibitemOpen
  \bibfield  {author} {\bibinfo {author} {\bibfnamefont {K.~W.}\ \bibnamefont
  {Schwarz}},\ }\bibfield  {title} {\bibinfo {title} {Vortex pinning in
  superfluid helium},\ }\href {https://doi.org/10.1103/physrevlett.47.251}
  {\bibfield  {journal} {\bibinfo  {journal} {Phys. Rev. Lett.}\ }\textbf
  {\bibinfo {volume} {47}},\ \bibinfo {pages} {251} (\bibinfo {year}
  {1981})}\BibitemShut {NoStop}%
\bibitem [{\citenamefont {Liu}\ \emph {et~al.}(2024)\citenamefont {Liu},
  \citenamefont {Prasad}, \citenamefont {Baggaley}, \citenamefont {Barenghi},\
  and\ \citenamefont {Wood}}]{Liu2024}%
  \BibitemOpen
  \bibfield  {author} {\bibinfo {author} {\bibfnamefont {I.-K.}\ \bibnamefont
  {Liu}}, \bibinfo {author} {\bibfnamefont {S.~B.}\ \bibnamefont {Prasad}},
  \bibinfo {author} {\bibfnamefont {A.~W.}\ \bibnamefont {Baggaley}}, \bibinfo
  {author} {\bibfnamefont {C.~F.}\ \bibnamefont {Barenghi}},\ and\ \bibinfo
  {author} {\bibfnamefont {T.~S.}\ \bibnamefont {Wood}},\ }\bibfield  {title}
  {\bibinfo {title} {Vortex depinning in a two-dimensional superfluid},\ }\href
  {https://doi.org/10.1007/s10909-024-03064-7} {\bibfield  {journal} {\bibinfo
  {journal} {J. Low Temp. Phys.}\ }\textbf {\bibinfo {volume} {215}},\ \bibinfo
  {pages} {376} (\bibinfo {year} {2024})}\BibitemShut {NoStop}%
\bibitem [{\citenamefont {Verma}\ \emph {et~al.}(2022)\citenamefont {Verma},
  \citenamefont {Pandit},\ and\ \citenamefont {Brachet}}]{Verma2022}%
  \BibitemOpen
  \bibfield  {author} {\bibinfo {author} {\bibfnamefont {A.~K.}\ \bibnamefont
  {Verma}}, \bibinfo {author} {\bibfnamefont {R.}~\bibnamefont {Pandit}},\ and\
  \bibinfo {author} {\bibfnamefont {M.~E.}\ \bibnamefont {Brachet}},\
  }\bibfield  {title} {\bibinfo {title} {Rotating self-gravitating
  {Bose-Einstein} condensates with a crust: {A} model for pulsar glitches},\
  }\href {https://doi.org/10.1103/PhysRevResearch.4.013026} {\bibfield
  {journal} {\bibinfo  {journal} {Phys. Rev. Res.}\ }\textbf {\bibinfo {volume}
  {4}},\ \bibinfo {eid} {013026} (\bibinfo {year} {2022})}\BibitemShut
  {NoStop}%
\bibitem [{\citenamefont {Warszawski}\ and\ \citenamefont
  {Melatos}(2011)}]{Warszawski2011}%
  \BibitemOpen
  \bibfield  {author} {\bibinfo {author} {\bibfnamefont {L.}~\bibnamefont
  {Warszawski}}\ and\ \bibinfo {author} {\bibfnamefont {A.}~\bibnamefont
  {Melatos}},\ }\bibfield  {title} {\bibinfo {title} {{Gross-Pitaevskii model
  of pulsar glitches}},\ }\href
  {https://doi.org/10.1111/j.1365-2966.2011.18803.x} {\bibfield  {journal}
  {\bibinfo  {journal} {Mon. Not. R. Astron. Soc.}\ }\textbf {\bibinfo {volume}
  {415}},\ \bibinfo {pages} {1611} (\bibinfo {year} {2011})}\BibitemShut
  {NoStop}%
\bibitem [{\citenamefont {Warszawski}\ \emph {et~al.}(2012)\citenamefont
  {Warszawski}, \citenamefont {Melatos},\ and\ \citenamefont
  {Berloff}}]{Warszawski2012}%
  \BibitemOpen
  \bibfield  {author} {\bibinfo {author} {\bibfnamefont {L.}~\bibnamefont
  {Warszawski}}, \bibinfo {author} {\bibfnamefont {A.}~\bibnamefont
  {Melatos}},\ and\ \bibinfo {author} {\bibfnamefont {N.~G.}\ \bibnamefont
  {Berloff}},\ }\bibfield  {title} {\bibinfo {title} {Unpinning triggers for
  superfluid vortex avalanches},\ }\href
  {https://doi.org/10.1103/physrevb.85.104503} {\bibfield  {journal} {\bibinfo
  {journal} {Phys. Rev. B}\ }\textbf {\bibinfo {volume} {85}},\ \bibinfo
  {pages} {104503} (\bibinfo {year} {2012})}\BibitemShut {NoStop}%
\bibitem [{\citenamefont {Melatos}\ \emph {et~al.}(2015)\citenamefont
  {Melatos}, \citenamefont {Douglass},\ and\ \citenamefont
  {Simula}}]{Melatos2015}%
  \BibitemOpen
  \bibfield  {author} {\bibinfo {author} {\bibfnamefont {A.}~\bibnamefont
  {Melatos}}, \bibinfo {author} {\bibfnamefont {J.~A.}\ \bibnamefont
  {Douglass}},\ and\ \bibinfo {author} {\bibfnamefont {T.~P.}\ \bibnamefont
  {Simula}},\ }\bibfield  {title} {\bibinfo {title} {Persistent gravitational
  radiation from glitching pulsars},\ }\href
  {https://doi.org/10.1088/0004-637x/807/2/132} {\bibfield  {journal} {\bibinfo
   {journal} {Astrophys. J.}\ }\textbf {\bibinfo {volume} {807}},\ \bibinfo
  {pages} {132} (\bibinfo {year} {2015})}\BibitemShut {NoStop}%
\bibitem [{\citenamefont {Shukla}\ \emph
  {et~al.}(2024{\natexlab{a}})\citenamefont {Shukla}, \citenamefont {Brachet},\
  and\ \citenamefont {Pandit}}]{Shukla2024}%
  \BibitemOpen
  \bibfield  {author} {\bibinfo {author} {\bibfnamefont {S.}~\bibnamefont
  {Shukla}}, \bibinfo {author} {\bibfnamefont {M.~E.}\ \bibnamefont
  {Brachet}},\ and\ \bibinfo {author} {\bibfnamefont {R.}~\bibnamefont
  {Pandit}},\ }\bibfield  {title} {\bibinfo {title} {Neutron-superfluid
  vortices and proton-superconductor flux tubes: Development of a minimal model
  for pulsar glitches},\ }\href {https://doi.org/10.1103/PhysRevD.110.083002}
  {\bibfield  {journal} {\bibinfo  {journal} {Phys. Rev. D}\ }\textbf {\bibinfo
  {volume} {110}},\ \bibinfo {pages} {083002} (\bibinfo {year}
  {2024}{\natexlab{a}})}\BibitemShut {NoStop}%
\bibitem [{\citenamefont {Lattimer}\ and\ \citenamefont
  {Prakash}(2004)}]{Lattimer2004}%
  \BibitemOpen
  \bibfield  {author} {\bibinfo {author} {\bibfnamefont {J.~M.}\ \bibnamefont
  {Lattimer}}\ and\ \bibinfo {author} {\bibfnamefont {M.}~\bibnamefont
  {Prakash}},\ }\bibfield  {title} {\bibinfo {title} {The physics of neutron
  stars},\ }\href {https://doi.org/10.1126/science.1090720} {\bibfield
  {journal} {\bibinfo  {journal} {Science}\ }\textbf {\bibinfo {volume}
  {304}},\ \bibinfo {pages} {536} (\bibinfo {year} {2004})}\BibitemShut
  {NoStop}%
\bibitem [{\citenamefont {Liu}\ \emph {et~al.}(2025)\citenamefont {Liu},
  \citenamefont {Baggaley}, \citenamefont {Barenghi},\ and\ \citenamefont
  {Wood}}]{Liu2025}%
  \BibitemOpen
  \bibfield  {author} {\bibinfo {author} {\bibfnamefont {I.-K.}\ \bibnamefont
  {Liu}}, \bibinfo {author} {\bibfnamefont {A.~W.}\ \bibnamefont {Baggaley}},
  \bibinfo {author} {\bibfnamefont {C.~F.}\ \bibnamefont {Barenghi}},\ and\
  \bibinfo {author} {\bibfnamefont {T.~S.}\ \bibnamefont {Wood}},\ }\bibfield
  {title} {\bibinfo {title} {Vortex avalanches and collective motion in neutron
  stars},\ }\href {https://doi.org/10.3847/1538-4357/adc383} {\bibfield
  {journal} {\bibinfo  {journal} {Astrophys. J.}\ }\textbf {\bibinfo {volume}
  {984}},\ \bibinfo {pages} {83} (\bibinfo {year} {2025})}\BibitemShut
  {NoStop}%
\bibitem [{\citenamefont {Nikolaieva}\ \emph {et~al.}(2023)\citenamefont
  {Nikolaieva}, \citenamefont {Bidasyuk}, \citenamefont {Korshynska},
  \citenamefont {Gorbar}, \citenamefont {Jia},\ and\ \citenamefont
  {Yakimenko}}]{Nikolaieva2023}%
  \BibitemOpen
  \bibfield  {author} {\bibinfo {author} {\bibfnamefont {Y.}~\bibnamefont
  {Nikolaieva}}, \bibinfo {author} {\bibfnamefont {Y.}~\bibnamefont
  {Bidasyuk}}, \bibinfo {author} {\bibfnamefont {K.}~\bibnamefont
  {Korshynska}}, \bibinfo {author} {\bibfnamefont {E.}~\bibnamefont {Gorbar}},
  \bibinfo {author} {\bibfnamefont {J.}~\bibnamefont {Jia}},\ and\ \bibinfo
  {author} {\bibfnamefont {A.}~\bibnamefont {Yakimenko}},\ }\bibfield  {title}
  {\bibinfo {title} {Stable vortex structures in colliding self-gravitating
  {Bose-Einstein} condensates},\ }\href
  {https://doi.org/10.1103/physrevd.108.023503} {\bibfield  {journal} {\bibinfo
   {journal} {Phys. Rev. D}\ }\textbf {\bibinfo {volume} {108}},\ \bibinfo
  {pages} {023503} (\bibinfo {year} {2023})}\BibitemShut {NoStop}%
\bibitem [{\citenamefont {Schiappacasse}\ and\ \citenamefont
  {Hertzberg}(2018)}]{Schiappacasse2018}%
  \BibitemOpen
  \bibfield  {author} {\bibinfo {author} {\bibfnamefont {E.~D.}\ \bibnamefont
  {Schiappacasse}}\ and\ \bibinfo {author} {\bibfnamefont {M.~P.}\ \bibnamefont
  {Hertzberg}},\ }\bibfield  {title} {\bibinfo {title} {Analysis of dark matter
  axion clumps with spherical symmetry},\ }\href
  {https://doi.org/10.1088/1475-7516/2018/01/037} {\bibfield  {journal}
  {\bibinfo  {journal} {J. Cosmol. Astropart. P.}\ }\textbf {\bibinfo {volume}
  {2018}},\ \bibinfo {pages} {037} (\bibinfo {year} {2018})}\BibitemShut
  {NoStop}%
\bibitem [{\citenamefont {Lee}\ \emph {et~al.}(2025)\citenamefont {Lee},
  \citenamefont {Kim}, \citenamefont {Jung},\ and\ \citenamefont
  {il~Shin}}]{Lee2025}%
  \BibitemOpen
  \bibfield  {author} {\bibinfo {author} {\bibfnamefont {J.}~\bibnamefont
  {Lee}}, \bibinfo {author} {\bibfnamefont {J.}~\bibnamefont {Kim}}, \bibinfo
  {author} {\bibfnamefont {J.}~\bibnamefont {Jung}},\ and\ \bibinfo {author}
  {\bibfnamefont {Y.}~\bibnamefont {il~Shin}},\ }\href
  {https://doi.org/10.48550/ARXIV.2502.07449} {\bibinfo {title} {Enhancement of
  damping in a turbulent atomic {Bose-Einstein} condensate}} (\bibinfo {year}
  {2025}),\ \Eprint {https://arxiv.org/abs/2502.07449} {arXiv:2502.07449
  [cond-mat.quant-gas]} \BibitemShut {NoStop}%
\bibitem [{\citenamefont {Ferrand}\ \emph {et~al.}(2021)\citenamefont
  {Ferrand}, \citenamefont {Sahraoui}, \citenamefont {Laveder}, \citenamefont
  {Passot}, \citenamefont {Sulem},\ and\ \citenamefont
  {Galtier}}]{Ferrand2021}%
  \BibitemOpen
  \bibfield  {author} {\bibinfo {author} {\bibfnamefont {R.}~\bibnamefont
  {Ferrand}}, \bibinfo {author} {\bibfnamefont {F.}~\bibnamefont {Sahraoui}},
  \bibinfo {author} {\bibfnamefont {D.}~\bibnamefont {Laveder}}, \bibinfo
  {author} {\bibfnamefont {T.}~\bibnamefont {Passot}}, \bibinfo {author}
  {\bibfnamefont {P.~L.}\ \bibnamefont {Sulem}},\ and\ \bibinfo {author}
  {\bibfnamefont {S.}~\bibnamefont {Galtier}},\ }\bibfield  {title} {\bibinfo
  {title} {Fluid energy cascade rate and kinetic damping: New insight from 3d
  landau-fluid simulations},\ }\href {https://doi.org/10.3847/1538-4357/ac2bfb}
  {\bibfield  {journal} {\bibinfo  {journal} {Astrophys. J.}\ }\textbf
  {\bibinfo {volume} {923}},\ \bibinfo {pages} {122} (\bibinfo {year}
  {2021})}\BibitemShut {NoStop}%
\bibitem [{\citenamefont {Shukla}\ \emph {et~al.}(2025)\citenamefont {Shukla},
  \citenamefont {Brachet},\ and\ \citenamefont
  {Pandit}}]{shukla2025selfgravitatingsuperfluidsgrosspitaevskiipoissonframework}%
  \BibitemOpen
  \bibfield  {author} {\bibinfo {author} {\bibfnamefont {S.}~\bibnamefont
  {Shukla}}, \bibinfo {author} {\bibfnamefont {M.~E.}\ \bibnamefont
  {Brachet}},\ and\ \bibinfo {author} {\bibfnamefont {R.}~\bibnamefont
  {Pandit}},\ }\href {https://arxiv.org/abs/2512.16936} {\bibinfo {title}
  {Self-gravitating superfluids: The {Gross-Pitaevskii-Poisson} framework}}
  (\bibinfo {year} {2025}),\ \Eprint {https://arxiv.org/abs/2512.16936}
  {arXiv:2512.16936 [astro-ph.HE]} \BibitemShut {NoStop}%
\bibitem [{\citenamefont {Sivakumar}\ \emph {et~al.}(2026)\citenamefont
  {Sivakumar}, \citenamefont {Mishra}, \citenamefont {Hujeirat},\ and\
  \citenamefont {Muruganandam}}]{Sivakumar2026}%
  \BibitemOpen
  \bibfield  {author} {\bibinfo {author} {\bibfnamefont {A.}~\bibnamefont
  {Sivakumar}}, \bibinfo {author} {\bibfnamefont {P.~K.}\ \bibnamefont
  {Mishra}}, \bibinfo {author} {\bibfnamefont {A.~A.}\ \bibnamefont
  {Hujeirat}},\ and\ \bibinfo {author} {\bibfnamefont {P.}~\bibnamefont
  {Muruganandam}},\ }\bibfield  {title} {\bibinfo {title} {Anomalous energy
  injection in the {Gross-Pitaevskii} framework for turbulence in neutron star
  glitches},\ }\href {https://doi.org/10.1103/3zdy-prmx} {\bibfield  {journal}
  {\bibinfo  {journal} {Phys. Rev. D}\ }\textbf {\bibinfo {volume} {113}},\
  \bibinfo {pages} {L041305} (\bibinfo {year} {2026})}\BibitemShut {NoStop}%
\bibitem [{\citenamefont {Shukla}\ \emph
  {et~al.}(2024{\natexlab{b}})\citenamefont {Shukla}, \citenamefont {Verma},
  \citenamefont {Brachet},\ and\ \citenamefont {Pandit}}]{Shukla2024Axion}%
  \BibitemOpen
  \bibfield  {author} {\bibinfo {author} {\bibfnamefont {S.}~\bibnamefont
  {Shukla}}, \bibinfo {author} {\bibfnamefont {A.~K.}\ \bibnamefont {Verma}},
  \bibinfo {author} {\bibfnamefont {M.~E.}\ \bibnamefont {Brachet}},\ and\
  \bibinfo {author} {\bibfnamefont {R.}~\bibnamefont {Pandit}},\ }\bibfield
  {title} {\bibinfo {title} {Gravity- and temperature-driven phase transitions
  in a model for collapsed axionic condensates},\ }\href
  {https://doi.org/10.1103/PhysRevD.109.063009} {\bibfield  {journal} {\bibinfo
   {journal} {Phys. Rev. D}\ }\textbf {\bibinfo {volume} {109}},\ \bibinfo
  {pages} {063009} (\bibinfo {year} {2024}{\natexlab{b}})}\BibitemShut
  {NoStop}%
\bibitem [{\citenamefont {Bradley}\ \emph {et~al.}(2022)\citenamefont
  {Bradley}, \citenamefont {Kumar}, \citenamefont {Pal},\ and\ \citenamefont
  {Yu}}]{Bradley2022}%
  \BibitemOpen
  \bibfield  {author} {\bibinfo {author} {\bibfnamefont {A.~S.}\ \bibnamefont
  {Bradley}}, \bibinfo {author} {\bibfnamefont {R.~K.}\ \bibnamefont {Kumar}},
  \bibinfo {author} {\bibfnamefont {S.}~\bibnamefont {Pal}},\ and\ \bibinfo
  {author} {\bibfnamefont {X.}~\bibnamefont {Yu}},\ }\bibfield  {title}
  {\bibinfo {title} {Spectral analysis for compressible quantum fluids},\
  }\href {https://doi.org/10.1103/physreva.106.043322} {\bibfield  {journal}
  {\bibinfo  {journal} {Phys. Rev. A}\ }\textbf {\bibinfo {volume} {106}},\
  \bibinfo {pages} {043322} (\bibinfo {year} {2022})}\BibitemShut {NoStop}%
\bibitem [{\citenamefont {Muruganandam}\ and\ \citenamefont
  {Adhikari}(2009)}]{Muruganandam2009}%
  \BibitemOpen
  \bibfield  {author} {\bibinfo {author} {\bibfnamefont {P.}~\bibnamefont
  {Muruganandam}}\ and\ \bibinfo {author} {\bibfnamefont {S.~K.}\ \bibnamefont
  {Adhikari}},\ }\bibfield  {title} {\bibinfo {title} {Fortran programs for the
  time-dependent {Gross-Pitaevskii} equation in a fully anisotropic trap},\
  }\href {https://doi.org/10.1016/j.cpc.2009.04.015} {\bibfield  {journal}
  {\bibinfo  {journal} {Comput. Phys. Commun.}\ }\textbf {\bibinfo {volume}
  {180}},\ \bibinfo {pages} {1888} (\bibinfo {year} {2009})}\BibitemShut
  {NoStop}%
\bibitem [{\citenamefont {Vudragovi{\'c}}\ \emph {et~al.}(2012)\citenamefont
  {Vudragovi{\'c}}, \citenamefont {Vidanovi{\'c}}, \citenamefont {Bala{\v{z}}},
  \citenamefont {Muruganandam},\ and\ \citenamefont
  {Adhikari}}]{Vudragovic2012}%
  \BibitemOpen
  \bibfield  {author} {\bibinfo {author} {\bibfnamefont {D.}~\bibnamefont
  {Vudragovi{\'c}}}, \bibinfo {author} {\bibfnamefont {I.}~\bibnamefont
  {Vidanovi{\'c}}}, \bibinfo {author} {\bibfnamefont {A.}~\bibnamefont
  {Bala{\v{z}}}}, \bibinfo {author} {\bibfnamefont {P.}~\bibnamefont
  {Muruganandam}},\ and\ \bibinfo {author} {\bibfnamefont {S.~K.}\ \bibnamefont
  {Adhikari}},\ }\bibfield  {title} {\bibinfo {title} {C programs for solving
  the time-dependent {Gross-Pitaevskii} equation in a fully anisotropic trap},\
  }\href {https://doi.org/10.1016/j.cpc.2012.03.022} {\bibfield  {journal}
  {\bibinfo  {journal} {Comput. Phys. Commun}\ }\textbf {\bibinfo {volume}
  {183}},\ \bibinfo {pages} {2021} (\bibinfo {year} {2012})}\BibitemShut
  {NoStop}%
\bibitem [{\citenamefont {Young-S}\ \emph {et~al.}(2017)\citenamefont
  {Young-S}, \citenamefont {Muruganandam}, \citenamefont {Adhikari},
  \citenamefont {Lon{\v{c}}ar}, \citenamefont {Vudragovi{\'c}},\ and\
  \citenamefont {Bala{\v{z}}}}]{YoungS2017}%
  \BibitemOpen
  \bibfield  {author} {\bibinfo {author} {\bibfnamefont {L.~E.}\ \bibnamefont
  {Young-S}}, \bibinfo {author} {\bibfnamefont {P.}~\bibnamefont
  {Muruganandam}}, \bibinfo {author} {\bibfnamefont {S.~K.}\ \bibnamefont
  {Adhikari}}, \bibinfo {author} {\bibfnamefont {V.}~\bibnamefont
  {Lon{\v{c}}ar}}, \bibinfo {author} {\bibfnamefont {D.}~\bibnamefont
  {Vudragovi{\'c}}},\ and\ \bibinfo {author} {\bibfnamefont {A.}~\bibnamefont
  {Bala{\v{z}}}},\ }\bibfield  {title} {\bibinfo {title} {{OpenMP GNU and Intel
  Fortran programs for solving the time-dependent {Gross-Pitaevskii}
  equation}},\ }\href {https://doi.org/10.1016/j.cpc.2017.07.013} {\bibfield
  {journal} {\bibinfo  {journal} {Comput. Phys. Commun}\ }\textbf {\bibinfo
  {volume} {220}},\ \bibinfo {pages} {503} (\bibinfo {year}
  {2017})}\BibitemShut {NoStop}%
\bibitem [{\citenamefont {Kumar}\ \emph {et~al.}(2019)\citenamefont {Kumar},
  \citenamefont {Lon{\v{c}}ar}, \citenamefont {Muruganandam}, \citenamefont
  {Adhikari},\ and\ \citenamefont {Bala{\v{z}}}}]{Kumar2019}%
  \BibitemOpen
  \bibfield  {author} {\bibinfo {author} {\bibfnamefont {R.~K.}\ \bibnamefont
  {Kumar}}, \bibinfo {author} {\bibfnamefont {V.}~\bibnamefont {Lon{\v{c}}ar}},
  \bibinfo {author} {\bibfnamefont {P.}~\bibnamefont {Muruganandam}}, \bibinfo
  {author} {\bibfnamefont {S.~K.}\ \bibnamefont {Adhikari}},\ and\ \bibinfo
  {author} {\bibfnamefont {A.}~\bibnamefont {Bala{\v{z}}}},\ }\bibfield
  {title} {\bibinfo {title} {{C and Fortran OpenMP} programs for rotating
  {Bose-Einstein} condensates},\ }\href
  {https://doi.org/10.1016/j.cpc.2019.03.004} {\bibfield  {journal} {\bibinfo
  {journal} {Comput. Phys. Commun}\ }\textbf {\bibinfo {volume} {240}},\
  \bibinfo {pages} {74} (\bibinfo {year} {2019})}\BibitemShut {NoStop}%
\bibitem [{\citenamefont {Lon\v{c}ar}\ \emph {et~al.}(2016)\citenamefont
  {Lon\v{c}ar}, \citenamefont {Bala\v{z}}, \citenamefont {Bogojevi\'{c}},
  \citenamefont {\v{S}krbi\'{c}}, \citenamefont {Muruganandam},\ and\
  \citenamefont {Adhikari}}]{Loncar2016}%
  \BibitemOpen
  \bibfield  {author} {\bibinfo {author} {\bibfnamefont {V.}~\bibnamefont
  {Lon\v{c}ar}}, \bibinfo {author} {\bibfnamefont {A.}~\bibnamefont
  {Bala\v{z}}}, \bibinfo {author} {\bibfnamefont {A.}~\bibnamefont
  {Bogojevi\'{c}}}, \bibinfo {author} {\bibfnamefont {S.}~\bibnamefont
  {\v{S}krbi\'{c}}}, \bibinfo {author} {\bibfnamefont {P.}~\bibnamefont
  {Muruganandam}},\ and\ \bibinfo {author} {\bibfnamefont {S.~K.}\ \bibnamefont
  {Adhikari}},\ }\bibfield  {title} {\bibinfo {title} {{CUDA} programs for
  solving the time-dependent dipolar {Gross-Pitaevskii} equation in an
  anisotropic trap},\ }\href {https://doi.org/10.1016/j.cpc.2015.11.014}
  {\bibfield  {journal} {\bibinfo  {journal} {Comput. Phys. Commun.}\ }\textbf
  {\bibinfo {volume} {200}},\ \bibinfo {pages} {406} (\bibinfo {year}
  {2016})}\BibitemShut {NoStop}%
\bibitem [{\citenamefont {Muruganandam}\ and\ \citenamefont
  {Radha}(2025)}]{Muruganandam2025}%
  \BibitemOpen
  \bibfield  {author} {\bibinfo {author} {\bibfnamefont {P.}~\bibnamefont
  {Muruganandam}}\ and\ \bibinfo {author} {\bibfnamefont {R.}~\bibnamefont
  {Radha}},\ }\href {https://doi.org/10.1088/978-0-7503-5447-9} {\emph
  {\bibinfo {title} {An Introduction to Ultracold Atoms with Analytical and
  Numerical Methods}}}\ (\bibinfo  {publisher} {IOP Publishing},\ \bibinfo
  {year} {2025})\BibitemShut {NoStop}%
\bibitem [{\citenamefont {Koonin}(2018)}]{Koonin2018}%
  \BibitemOpen
  \bibfield  {author} {\bibinfo {author} {\bibfnamefont {S.~E.}\ \bibnamefont
  {Koonin}},\ }\href {https://doi.org/https://doi.org/10.1201/9780429494024}
  {\emph {\bibinfo {title} {Computational Physics}}}\ (\bibinfo  {publisher}
  {Chapman and Hall/CRC},\ \bibinfo {address} {Boulder},\ \bibinfo {year}
  {2018})\BibitemShut {NoStop}%
\end{thebibliography}
\end{document}